\documentclass[10pt,journal,final,twocolumn,]{IEEEtran}
\usepackage{amsmath,amssymb,amsfonts}
\usepackage{caption} 
\usepackage{bm}
\usepackage{algorithmicx}
\usepackage{algorithm}
\usepackage{graphicx}
\usepackage{float}
\usepackage{textcomp}
\usepackage{xcolor}
\usepackage{subfigure}
\usepackage[noend]{algpseudocode}
\usepackage[numbers,sort&compress]{natbib}
\usepackage{mathrsfs}
\usepackage{verbatim}
\usepackage{balance}
\usepackage{booktabs}
\usepackage{fancybox}
\usepackage{changepage}
\usepackage{extarrows}
\usepackage{diagbox}
\usepackage{multirow}
\usepackage{array}
\usepackage{epstopdf}
\usepackage{soul}

\newtheorem{lemma}{Lemma}

\def\BibTeX{{\rm B\kern-.05em{\sc i\kern-.025em b}\kern-.08em
		T\kern-.1667em\lower.7ex\hbox{E}\kern-.125emX}}

\newcommand{\bE}{{\bf E}}
\newcommand{\bw}{{\bf w}}

\newcommand{\bT}{{\bf T}}

\newcommand{\bZ}{{\bf Z}}

\begin{document}
	\title{Reconfigurable Intelligent Surface-Enabled Green and Secure Offloading for Mobile Edge Computing Networks\\}
	\author{\IEEEauthorblockN{Tong-Xing Zheng, \emph{Senior Member, IEEE}, Xinji Wang, Xin Chen, Di Mao, Jia Shi, \emph{Member, IEEE}, Cunhua Pan, \emph{Senior Member, IEEE}, Chongwen Huang, \emph{Member, IEEE}, Haiyang Ding,  \emph{Member, IEEE}, and Zan Li, \emph{Fellow, IEEE}}
        \thanks{Tong-Xing Zheng is with the School of Information and Communications Engineering, Xi'an Jiaotong University, Xi'an 710049, China, with the National Key Laboratory of Multi-domain Data Collaborative Processing and Control, Xi'an 710068, China, and also with the Ministry of Education Key Laboratory for Intelligent Networks and Network Security, Xi'an Jiaotong University, Xi'an 710049, China (e-mail: zhengtx@mail.xjtu.edu.cn).}
	\thanks{
			Xinji Wang is with the Huawei Technologies Co., Ltd., Shanghai 200040, China (e-mail: icebag@stu.xjtu.edu.cn).}
		\thanks{Xin Chen is with the Purple Mountain Laboratories, Nanjing 211111, China (e-mail: chenxin02@pmlabs.com.cn).}
		\thanks{Di Mao is with the National Key Laboratory of Multi-domain Data Collaborative Processing and Control, Xi'an 710068, China (e-mail: mdcom1@126.com).}
        \thanks{Jia Shi and Zan Li are with the State Key Laboratory of Integrated Service Networks, Xidian University, Xi'an, 710071, China, and also with the School of Telecommunications Engineering, Xidian University, Xi'an, 710071, China (e-mail: jiashi@xidian.edu.cn; zanli@xidian.edu.cn).}
	\thanks{Cunhua Pan is with the National Mobile Communications Research Laboratory, Southeast University, Nanjing 211111, China (e-mail: cpan@seu.edu.cn).}
	\thanks{Haiyang Ding is with the School of Information and Communications, National University of Defense Technology, Wuhan 430035, China (e-mail: dinghy2003@nudt.edu.cn).}
	\thanks{Chongwen Huang is with the College of Information Science and Electronic Engineering, Zhejiang University, Hangzhou 310027, China, and also with Zhejiang Provincial Key Laboratory of Info. Proc., Commun. \& Netw., Hangzhou 310027, China (e-mail: chongwenhuang@zju.edu.cn).}
           }

	\maketitle
	{ 
	\begin{abstract}
    This paper investigates a multi-user uplink mobile edge computing (MEC) network, where the users offload partial tasks securely to an access point under the non-orthogonal multiple access policy with the aid of a reconfigurable intelligent surface (RIS) against a multi-antenna eavesdropper. We formulate a non-convex optimization problem of minimizing the total energy consumption subject to secure offloading requirement, and we build an efficient block coordinate descent framework to iteratively optimize the number of local computation bits and transmit power at the users, the RIS phase shifts, and the multi-user detection matrix at the access point. Specifically, we successively adopt successive convex approximation, semi-definite programming, and semidefinite relaxation to solve the problem with perfect eavesdropper's channel state information (CSI), and we then employ S-procedure and penalty convex-concave to achieve robust design for the imperfect CSI case. We provide extensive numerical results to validate the convergence and effectiveness of the proposed algorithms. We demonstrate that RIS plays a significant role in realizing a secure and energy-efficient MEC network, and deploying a well-designed RIS can save energy consumption by up to 60\% compared to that without RIS. We further reveal impacts of various key factors on the secrecy energy efficiency, including RIS element number and deployment position, user number, task scale and duration, and CSI imperfection.       
	\end{abstract}
}

    \section{Introduction}
{ With the evolution of wireless networks from connected things to connected intelligence, mobile edge computing (MEC) has been increasingly recognized as an appealing approach for providing efficient and intelligent services. 
Different from conventional cloud computing, MEC offloads computation tasks from core data center to network edge server which is much closer to end users, thus significantly improving computation capacity and response speed for wireless networks \cite{Mao2017Survey}.
Despite the advantages, the inherent broadcast nature of electromagnetic signal propagation renders task offloading vulnerable to eavesdropping attacks, which necessitates a targeted consideration of security solutions when designing an MEC network. Physical layer security (PLS) \cite{Wang2016Physical,Zheng2022Physical,Lin2025Wireless}, an essential security mechanism beyond traditional encryption methods, guarantees wireless transmission secrecy by leveraging the characteristics of wireless channels and noise randomness, and has a great potential in security reinforcement for MEC networks. 

Nevertheless, PLS is generally channel-dependent, which implies the achievable security performance might be heavily deteriorated under poor channel conditions. Fortunately, reconfigurable intelligent surfaces (RIS) emerges as a transformative technology to break through such restriction via paradigm shift from channel adaptation to channel reconstruction, thus offering new degrees of freedom to wireless communication systems \cite{Tang2020MIMO,Wu2021Intelligent,Li2022How,Pan2022Overview}. Specifically, RIS is composed of a large number of low-cost reflecting elements each of which can adjust the amplitude and phase of incident signals independently and controllably, and in this way we can reconstruct wireless propagation environment and create channel advantages for realizing PLS \cite{Shen2019Secrecy,Cui2019Secure,Li2022Intelligent,Lin2024Self,Dong2025Secure,Hong2020ArtificialNoiseAided}. 

Motivated by the various advantages of PLS and RIS, this paper aims to exploit the potential of a joint utilization of them for an MEC network, with an emphasize on the realization of an energy-efficient and secure multi-user MEC network even suffering from imperfect channel state information (CSI).
}

\subsection{Related Works}
{ In the context of PLS, plenty of research has already been carried out on the secure offloading for MEC networks. 
For instance, a secure multi-carrier transmission method is proposed in  \cite{Xu2019Exploiting} to safeguard multi-user task offloading against eavesdroppers. 
The joint design of task allocation, CPU frequency, transmit power, and offloading scheduling is studied in \cite{Wang2020Joint} to minimize the total energy consumption under secure offloading requirements. 
The integration of UAV-assisted cooperative jamming is investigated in \cite{Zhou2020Secure} to maximize the minimum secrecy capacity for an MEC-UAV network.

In order to improve the spectrum efficiency of multi-user task offloading, the non-orthogonal multiple access (NOMA) technique has been introduced into an MEC network, which allows users to share the same spectrum resources to transfer their data to the edge server simultaneously.
The PLS issues have also been examined for the NOMA-MEC networks. Specifically, a secure MEC scheme is proposed in \cite{Qian2021SecrecyBased}, where conventional devices collaborate with edge devices in an NOMA transmission group to assist in jamming eavesdroppers. 
The anti-eavesdropping capability is maximized in \cite{Wu2021Resource} for an uplink NOMA system. A cooperative jamming technique is leveraged in \cite{Wu2022NonOrthogonal} to enhance the secrecy performance of computation offloading. The total energy consumption is minimized in \cite{Zheng2024Secure} for a secure NOMA-MEC system by adopting the secrecy outage probability as a secrecy performance metric. On the other hand, the benefit of deploying an RIS has further been discussed for the security-oriented NOMA networks. 
For example, robust secure transmission is studied in \cite{Zhang2021Robust} for an RIS-assisted NOMA with eavesdropper's imperfect CSI, where the transmit power is minimized via a joint optimization of transmit beamforming and RIS phase shifts. 
This work is further extended to the robust secure NOMA system with the aid of distributed RISs, where the minimum secrecy rate among users is maximized by jointly optimizing transmit beamforming and RIS phase shifts leveraging only the eavesdropper's statistical CSI \cite{Zhang2022Securing}.  
The heterogeneous internal secrecy requirements are further addressed in  \cite{Li2021Intelligent} for the RIS-aided downlink transmission.

The potential of MEC cannot be fully
utilized when the communication links for task offloading
are hostile. Therefore, researchers have begun exploring the advantage of integrating the RIS into an MEC network to enhance the efficiency of task offloading efficiency. 
To be more specific, an RIS-assisted MEC is investigated in \cite{Bai2020Latency}, where the system latency is minimized by jointly optimizing task resource allocation, multi-user detection (MUD) matrix, and RIS phase shifts. This model is extended to an RIS-enabled NOMA-MEC network, where simultaneous wireless information and power transfer is considered \cite{Yang2024Delay}. A novel simultaneously transmitting and reflecting RIS (STAR-RIS) structure is proposed in \cite{Zhang2023Resource} to further improve the MEC performance. An MEC-enabled Internet of Things (IoT) network coexisting with both aerial and ground eavesdroppers is considered in a recent work \cite{Michailidis2024Optimization}, where an RIS-aided security-aware task offloading framework is proposed under the time-division multiple access (TDMA) protocol to improve the secure computation efficiency.} 

\begin{table*}[t]
	{ \centering
		\caption{\textsc{Comparison between our work and related works}}
		\label{tab:relatedwork}
		\begin{tabular}{|c|c|c|c|c|c|c|c|}
			\hline
			\textbf{Reference} & \textbf{Research Topic} & \textbf{Design Objective} & \textbf{PLS} & \textbf{RIS} & \textbf{MEC} &  \textbf{NOMA} &  \textbf{Imperfect CSI} \\ \hline
			\cite{Xu2019Exploiting} & Secure multi-user multi-carrier offloading & Minimize total energy consumption & \checkmark & $\times$ & \checkmark & $\times$ & \checkmark \\ \hline
			\cite{Wang2020Joint} & Secure multi-user offloading & Minimize total energy consumption & \checkmark & $\times$ & \checkmark & $\times$& $\times$ \\ \hline
			\cite{Zhou2020Secure} & Secure full-duplex UAV-MEC & Maximize minimum secrecy capacity & \checkmark & $\times$ & \checkmark & $\times$ & $\times$ \\ \hline
			\cite{Qian2021SecrecyBased} & Cooperative NOMA for secure MEC & Minimize
			total energy consumption & \checkmark & $\times$ & \checkmark & \checkmark & $\times$\\ \hline
			\cite{Wu2021Resource} & Two-user uplink NOMA for secure MEC & Maximize anti-eavesdropping ability & \checkmark & $\times$ & \checkmark & \checkmark & \checkmark\\ \hline
			\cite{Wu2022NonOrthogonal} & Secure offloading via cooperative jamming  &  Minimize total energy consumption & \checkmark & $\times$ & \checkmark & \checkmark & \checkmark\\ \hline
			\cite{Zheng2024Secure} & Secure and green multi-user offloading& Minimize total energy consumption& \checkmark & $\times$ & \checkmark & \checkmark & \checkmark\\ \hline
			\cite{Zhang2021Robust} & RIS-assisted two-user secure NOMA & Minimize transmit power & \checkmark  & \checkmark & $\times$ & \checkmark & \checkmark\\ \hline
			\cite{Zhang2022Securing} & Secure NOMA via distributed RISs& Maximize minimum secrecy rate & \checkmark & \checkmark & $\times$ & \checkmark & \checkmark \\ \hline
			\cite{Li2021Intelligent} & Internal secrecy in downlink NOMA & Minimize total transmit power & \checkmark & \checkmark & $\times$ & \checkmark & $\times$ \\ \hline
			\cite{Bai2020Latency} & RIS-aided uplink multi-user MEC & Minimize weighted delay & $\times$ & \checkmark & \checkmark & $\times$ & $\times$\\ \hline
			\cite{Yang2024Delay} & RIS-aided NOMA-MEC with SWIPT & Minimize delay \& energy consumption & $\times$ & \checkmark & \checkmark & \checkmark & $\times$\\ \hline
			\cite{Zhang2023Resource} & STAR-RIS-aided NOMA-MEC &  Minimize total energy consumption & $\times$ & \checkmark & \checkmark & \checkmark & $\times$\\ \hline
			\cite{Michailidis2024Optimization} & UAV-Enabled RIS-aided MEC-IoT & Maximize secure computation efficiency & \checkmark & \checkmark & \checkmark & $\times$ & \checkmark \\ \hline
			Our work & RIS-aided secure and green NOMA-MEC & Minimize total energy consumption & \checkmark & \checkmark & \checkmark & \checkmark & \checkmark \\ \hline
		\end{tabular}}
	\end{table*}
	
\subsection{Motivations and Contributions}
{ Table~\ref{tab:relatedwork} provides a comparative summary of the above-mentioned related works.} It is worth noting that PLS and RIS have been well investigated separately for MEC networks, while their combinative advantage has not yet been fully exploited. 
Moreover, the secrecy performance and energy efficiency of the existing computation offloading framework will deteriorate sharply when the CSI of eavesdropper is not available. 
{ On the other hand, recent developments in RIS prototypes and field trials have strengthened the feasibility of deploying RIS in real-world wireless systems, paving the way for integrating RIS into MEC networks \cite{Tang2020MIMO,Pan2021Reconfigurable}.} Motivated by all these, this paper investigates a novel RIS-assisted multi-user uplink MEC network under the monitoring of a multi-antenna eavesdropper, and we design both computation resource allocation and task offloading scheme to realize a secure and energy-efficient MEC network, considering both scenarios of perfect and imperfect eavesdropper's CSI. Specifically, the main contributions are summarized as follows:
\begin{itemize}
{ 
\item 
We establish a new research paradigm to combine the PLS and RIS technologies organically into a multi-user MEC network.
We propose a hybrid communication and computation MEC architecture, which executes users' partial computation tasks locally while offloading the rest to a multi-antenna access point (AP) for edge computing, leveraging the RIS-aided secure NOMA transmission. 
\item
We build a comprehensive communication and computation optimization framework to minimize the total energy consumption for local computing and edge offloading while satisfying the secure offloading requirement. We formulate a high-dimension multivariate optimization problem of energy consumption minimization for both perfect and imperfect CSI cases, and propose a joint design scheme of computation resource allocation, user transmit power, MUD matrix, as well as RIS phase shifts. 
\item 
We develop a block coordinate descent (BCD) algorithm to efficiently tackle the formulated complicated non-convex optimization problem. We divide the optimization variables into three blocks, namely, the number of local computation bits and transmit power at the users, the phase shifts at the RIS, and the MUD matrix at the AP. We successively adopt successive convex approximation (SCA), semi-definite programming (SDP), and semidefinite relaxation (SDR) to derive the optimal solution for the perfect CSI case, and we further employ S-procedure and penalty convex-concave (PCC) methods to achieve robust design for the imperfect CSI case.   
\item 
We provide extensive numerical results to validate the convergence and effectiveness of the proposed design methods. We also gain various insights into the design of the RIS-aided MEC networks, including deploying a moderate number of RIS elements to balance well between enhanced secrecy energy efficiency and increased system overhead, and how the RIS position, the duration for task offloading, the number of computation bits at each user, as well as the imperfection of eavesdropper's CSI will influence the system performance. 
}
\end{itemize}

	\subsection{Organization and Notation}
	The rest of the paper is organized as follows. 
	Section II describes the system model. 
	In Sections III and IV, we deal with the optimization problem for both cases of perfect and imperfect eavesdropper's CSI. 
	Numerical results are presented in Section V. 
	Section VI concludes this paper.
	
	\emph{Notations}: Bold uppercase and lowercase letters denote matrices and vectors, respectively.
	 $|\cdot|$, $\|\cdot\|$, $(\cdot)^{\dagger}$, $(\cdot)^{\rm T}$, $(\cdot)^{\rm H}$, $\ln(\cdot)$, and $\mathbb{E}\{\cdot\}$ denote the absolute value, Euclidean norm, conjugate, transpose, Hermitian transpose, natural logarithm, and expectation, respectively. 
	$ \mathbb{C}^{M\times N} $ denotes the $ M\times N $ complex-valued matrix space. 
	$ \mathrm{Re}(a) $ denotes the real part of scalar $ a $. 
	$ \mathrm{diag}\left( \mathbf{a} \right) $ denotes a diagonal matrix with all diagonal entries from vector $\mathbf{a} $.  
	$ \mathbf{A} \succeq 0 $ means that matrix $\mathbf{A}$ is positive semi-definite. $ \mathrm{Tr}(\mathbf{A}) $ and $ \mathrm{rank}(\mathbf{A}) $ denote the trace and rank of matrix $ \mathbf{A} $, respectively. $[x]^+\triangleq \max(x,0)$.
	\section{System Model}
	
		\begin{figure}[!t]
			\centering
			\includegraphics[width=\linewidth]{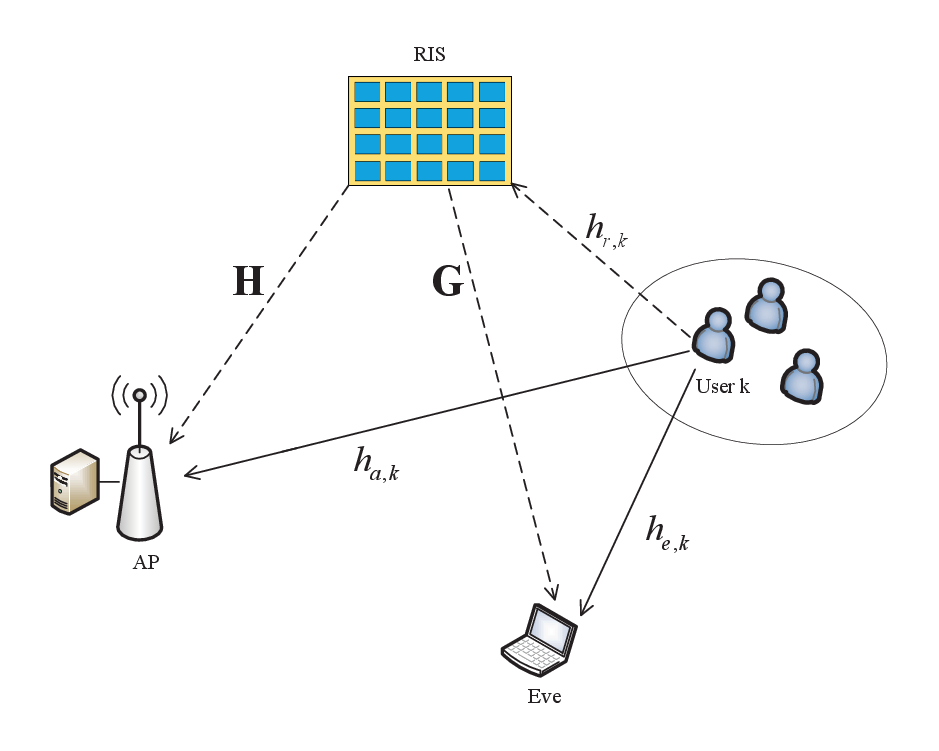} 
			\caption{Illustration of an RIS-assisted uplink MEC network with secure offloading from $ K $ users to an AP against an Eve.}	
			\label{systemmodelno}
		\end{figure}
	{ We consider an RIS-assisted MEC network as depicted in Fig. \ref{systemmodelno}, where $K$ single-antenna users simultaneously offload their partial computation tasks to an $N_a$-antenna AP integrated with an MEC server on the same frequency band, coexisting with an $N_e$-antenna eavesdropper (Eve) who attempts to intercept the offloaded information bits. An RIS comprised of $M$ reflecting elements is introduced to facilitate a secure and green MEC network in two ways. First, by adjusting the phase shifts of incident signals controllably, the RIS can reconfigure the wireless propagation environment to enhance the strength of AP's received signal while suppressing that of Eve's, thus improving offloading security. Second, owing to passive beamforming of RIS, users are allowed to offload tasks with lower transmit power, which reduces energy consumption of maintaining the same offloading rate.}	

	We consider a quasi-static flat-fading channel model, and the AP antenna spacing and RIS element spacing are large enough to ensure channel fading independence between antennas and between RIS elements. The channel coefficients from user $ k $ to AP, to RIS, and to Eve are denoted as $ \mathbf{h}_{a,k}\in \mathbb{C}^{N_a\times1} $, $ \mathbf{h}_{r,k}\in \mathbb{C}^{M\times1} $, and $ \mathbf{h}_{e,k}\in \mathbb{C}^{N_e\times1} $, respectively, $ k \in\mathcal{K} \triangleq \left\{1, \cdots, K\right\}$. The channels spanning from RIS to Eve and to AP are denoted by $ \mathbf{G}\in\mathbb{C}^{N_e\times M} $ and $ \mathbf{H}\in\mathbb{C}^{N_a\times M} $, respectively. The RIS phase shifts are characterized by a diagonal matrix $ \mathbf{\Theta}\triangleq \mathrm{diag}\left(\left[e^{j\theta_1},e^{j\theta_2},\cdots,e^{j\theta_M}\right]\right) $, where $ \theta_m \in \left[0, 2\pi\right) $ denotes the phase shift of the $ m $-th reflecting element, $ m\in\mathcal{M} \triangleq \left\{1, \cdots, M\right\} $. { We adopt a common assumption that the CSI of legitimate channels $\mathbf{h}_{a,k}$, $\mathbf{h}_{r,k}$, and $\mathbf{H}$ can be perfectly acquired at AP through  periodical channel estimation methods \cite{He2021Channel}. As for Eve's channels $\mathbf{h}_{e,k}$ and $\mathbf{G}$, we will consider both perfect and imperfect CSI scenarios, given that exact channel estimation for a non-cooperative and passive eavesdropper is generally challenging in practice.}
	
	\subsection{Secure Offloading}
	In the MEC system, each user $ k\in \mathcal{K} $ must execute a computation task of $ L_k $ bits within duration $T$, with a portion of its computation tasks, i.e., $ l_k <L_k $ bits, being accomplished locally and the remainder $ L_k-l_k $ bits being offloaded to the AP. The NOMA technique is adopted to enhance spectral efficiency, which enables $ K $ users to communicate with the AP simultaneously \cite{Liu2017Nonorthogonal}.\footnote{{Pure NOMA might cause larger delay or energy consumption due to higher complexity than hybrid NOMA and OMA like TDMA or FDMA, but such performance loss can be compensated for via parameter optimization. Besides, pure NOMA can be readily extended to hybrid NOMA which divides users into different groups with OMA and NOMA deployed for inter- and intra-group, respectively.
			Thus, our obtained results can be directly applied for inter-group NOMA, while inter-group OMA is much easier to realize, only requiring orthogonal time and frequency resource blocks.}} Thus, the received signals at the AP and the Eve can be represented by
	\begin{align}
		\label{key}
		\mathbf{y}_a &= \sum_{k=1}^K \sqrt{p_k}{\mathbf{h}_{k}}s_k +\mathbf{n}_a,\\
		\label{key2}
		\mathbf{y}_e &= \sum_{k=1}^K \sqrt{p_k}{\mathbf{g}_{k}}s_k +\mathbf{n}_e,
	\end{align}
	where $ \mathbf{h}_{k} \triangleq  \mathbf{h}_{a,k}+\mathbf{H}\mathbf{\Theta}\mathbf{h}_{r,k}$ and $ \mathbf{g}_{k} \triangleq \mathbf{h}_{e,k}+\mathbf{G}\mathbf{\Theta}\mathbf{h}_{r,k} $ with $ k \in\mathcal{K}$, denote the equivalent channels from user $ k $ to the AP and Eve, respectively. $ s_k $ and $ p_k $ are the information symbol and the transmit power of user $ k $, respectively, and $ \mathbf{n}_a \in \mathbb{C}^{N_a\times1} $ and $ \mathbf{n}_e \in \mathbb{C}^{N_e\times1} $ denote the noise at the AP and Eve, respectively. Without loss of generality, we assume that $ \mathbb{E}\left\{\left|s_{k}\right|\right\}=1 $, $ \mathbb{E}\left\{\mathbf{n}_a \mathbf{n}_a^{\rm H}\right\}=\sigma_a^{2} \mathbf{I}_{N_a} $, and $ \mathbb{E}\left\{\mathbf{n}_e \mathbf{n}_e^{\rm H}\right\}=\sigma_e^{2} \mathbf{I}_{N_e} $.

    For the uplink NOMA policy, the AP employs the SIC technique to decode the received signals \cite{Zheng2024Secure}.\footnote{{Here we use SIC as the multi-user decoding method mainly due to its ease of implementation and low computational complexity, which is well-suited for real-time and resource-constrained environments such as MEC.}} In general, the SIC decoding order of NOMA should obey a decreasing order of transmitter-to-receiver channel gains.
    However, if we determine the decoding order solely based on the direct channels from users to AP $\mathbf{h}_{a,k}$ while ignoring the reflecting channels from RIS to AP, we might suffer from severe performance loss, since the channel reconfiguration brought by RIS would probably change the channel gain order significantly. Instead, it would be more beneficial to take the SIC decoding order as a descending order of the equivalent channel gains $\left\|\mathbf{h}_{k}\right\|$ among users. 
     Without loss of generality, we assume that $ K $ users are sorted in a descending order of $\left\|\mathbf{h}_{k}\right\|$ such that $\left\|\mathbf{h}_{1}\right\|>\left\|\mathbf{h}_{2}\right\|>\cdots>\left\|\mathbf{h}_{K}\right\| $. 
     It is worth noting that, for the optimization problem examined in subsequent sections, since $\mathbf{h}_k $ depends on the phase shift matrix $ \mathbf{\Theta} $, the SIC decoding order should be updated along with the iteration of $ \mathbf{\Theta} $.

    We assume that AP utilizes detection vector $ \mathbf{w}_k\in \mathbb{C}^{N_a\times 1} $ with $ \left\| \mathbf{w}_k \right\|^2 = 1 $ to recover the signal of user $ k $. { We also assume that perfect SIC can be performed for uplink NOMA transmission, such that when decoding the signal from user $k$, the interfering signals from stronger users $j<k$ can be decoded and canceled completed while only leaving the signals from weaker users $j>k$ as interference.\footnote{{Imperfect SIC might introduce residual interference and reduce secrecy rate, but such rate loss could be compensated by posing a more tight secrecy rate constraint, while the subsequent problem structure and optimization framework would remain unchanged essentially. The perfect SIC assumption, in contrast, makes it much more convenient to tackle complicated parameter optimization and to gain useful insights into system design and analysis.}}} Thus, the signal-to-interference-plus-noise ratio (SINR) of user $ k $ can be calculated as    
	\begin{equation}
		\label{user_SINR}
		\gamma_k = \frac{p_k\left|\mathbf{w}_k^{\rm H}\mathbf{h}_k\right|^2}{\sum_{j=k+1}^{K}{p_j\left|\mathbf{w}_k^{\rm H}\mathbf{h}_j\right|^2}+\sigma_a^{2} \left\|\mathbf{w}_k \right\|^2 },\quad\forall k\in \mathcal{K}.
	\end{equation}
    	
		From a robust design perspective, we consider a worst-case scenario where Eve has powerful ability of parallel interference cancellation such that all the inter-user interference can be eliminated. 
		Note that although this assumption might overestimate Eve's decoding capability, it provides an upper bound on the total energy consumption of a secure and green MEC network and ensures effectiveness of our design results even for various less severe eavesdropping scenarios \cite{Zheng2024Secure}. In this case, Eve's received SINR for decoding the signal from user $ k $ can be given by
	\begin{equation}
	\label{eve_SINR}
		\gamma_{e,k}=\frac{p_k\left\|\mathbf{g}_{k}\right\|^2}{\sigma_e^{2} },\quad\forall k\in \mathcal{K}.
	\end{equation}
    
	 By overestimating Eve's wiretap capability, a lower bound on the achievable secrecy rate of user $k$ can be given by { \cite{Wang2016Physical}}
\begin{equation}
	\label{ach_rate}
	R_{s,k} = \left[\log_2\left({1+\gamma_k}\right)-\log_2\left({1+\gamma_{e,k}}\right)\right]^+,\quad\forall k\in \mathcal{K}.
\end{equation}
	
	In addition, the energy consumption of user $ k $ for computation offloading is given by
	\begin{equation}
		E_k^{\rm off} = p_kT, \quad\forall k\in \mathcal{K}.
	\end{equation}
	\subsection{Local Computing}
	For local computing, let \( C_k \) denote the number of CPU cycles required for computing one input bit at user \( k \). Then, the total number of CPU cycles required for computing \( l_k \) bits at user \( k \) is \( C_k l_k \). For the sake of energy efficiency, for each CPU cycle \( n \in \{ 1, \cdots, C_k l_k \} \), user \( k \) can control the CPU frequency \( f_{k,n} \) by employing the dynamic voltage and frequency scaling technique. As a consequence, the total execution time for local computing of user \( k \) is \( \sum_{n=1}^{C_k l_k} {f_{k,n}^{-1}} \). To ensure local computing to be accomplished within duration \( T \), the local execution time should satisfy \( \sum_{n=1}^{C_k l_k} {f_{k,n}^{-1}}  \le T \). The energy consumption of user \( k \) for local computing is \( E_k^{\rm loc} = \sum_{n=1}^{C_k l_k} \varsigma_k f_{k,n}^{2} \), where \( \varsigma_k > 0 \) is the effective capacitance coefficient of the chip architecture for user \( k \).
	
	Since the above time constraint and energy consumption are convex with CPU frequencies \( f_{k,n} \), the optimal solution for minimizing local energy consumption while ensuring computation latency \( T \) is to have identical CPU frequencies \( f_{k,n} \) over different cycles \cite{Fang2020Optimal}. As a result, the CPU frequencies at user \( k \) should be chosen as \( f_{k,n} = {C_k l_k}/{T} \) for \( n \in \{ 1, \cdots, C_k l_k \} \). The corresponding energy consumption can be calculated as
	\begin{equation}
		E_k^{\rm loc}=\frac{{{\varsigma }_{k}}{C_k^3}{l_k^3}}{T^2},\quad\forall k\in \mathcal{K}.
	\end{equation}
	
	The energy consumption for user $k$ consists of two components: local computing energy $ E_k^{\rm loc} $ and offloading energy $ E_k^{\rm off} $. Hence, the total energy consumption of $K$ users is 
	\begin{equation}
    \label{energy_equation}
		E = \underset{k\in \mathcal{K}}{\sum}\left(\frac{{{\varsigma }_k}{C_k^3}{l_k^3}}{T^2}+{p_kT}\right).
	\end{equation}
	
	We aim at minimizing the total energy consumption $E$ while guaranteeing a certain level of offloading secrecy. In the following section, we first deal with the total energy consumption minimization problem considering the cases of perfect and imperfect Eve's CSI to gain useful insights to system design and performance analysis. For each case, we will jointly design the optimal solution on computing resource allocation, user transmit power, AP MUD matrix, and RIS phase shifts to achieve our goal.
	
	\section{Secure Offloading with Perfect CSI}
	In this section, the total energy consumption minimization problem with perfect CSI of Eve is studied. The performance achieved under the perfect CSI case can serve as an upper bound for the proposed secure offloading design.
	\subsection{Problem Formulation}
	As discussed in \cite{Wang2021Energy}, Eves may have previously been legitimate users but are no longer authorized to access confidential information in the current communication process, or the AP has decided not to transmit such information to them. To ensure communication security, the AP treats these users as potential eavesdroppers. Since these eavesdroppers remain part of the communication system as receivers, the AP can obtain their CSI accurately through channel estimation and feedback. 
	
	Under the perfect CSI model, the total energy consumption minimization problem is formulated as
    { 
	\begin{subequations}
		\label{optimal problem}
		\begin{align}
			\underset{\mathbf{l},\mathbf{p},\mathbf{W},\mathbf{\Theta}}{\mathop{\mathrm{min}}}
		~&E = \underset{k\in \mathcal{K}}{\sum}\left(\frac{{{\varsigma}_k}{C_k^3}{l_k^3}}{T^2}+{p_kT}\right),\\
			\mathrm{s.t.} ~
			\label{sec_rate}
			&BTR_{s,k} \ge L_k-l_k,~\forall k\in \mathcal{K},\\
			\label{power}	
			&0\le p_k \le P_k,~\forall k\in \mathcal{K},\\
			\label{bits}
			& 0\le l_k \le L_k,~\forall k\in \mathcal{K},\\
			\label{shift}
			&\left|e^{j\theta_m}\right|=1,~\forall m\in \mathcal{M},
		\end{align}
	\end{subequations}
	}where $\mathbf{l}\triangleq[{l}_{1},\cdots,{{l}_{K}}]$ denotes the vector of local computing bits, $\mathbf{p}\triangleq[{p}_{1},\cdots,{{p}_{K}}]$ denotes the transmit power vector, $\mathbf{W} \triangleq \left[\bw_1, \dots, \bw_K\right]$ denotes the MUD matrix, and $\mathbf{\Theta}$ denotes the phase shift matrix. {  In the above problem,  \eqref{sec_rate} represents a secrecy rate constraint, which guarantees that the confidential computation bits can be securely and reliably offloaded to AP with limited duration $T$ and bandwidth $B$;}
	\eqref{power} represents a transmit power constraint, which ensures that each user's transmit power $p_ k $ should not exceed its power budget $P_k$; \eqref{bits} represents a local computation constraint; \eqref{shift} represents a unit modulus constraint (UMC), which indicates each reflecting element of RIS only adjusts the phases of the incident signals.
		
		{  Note that Problem \eqref{optimal problem} is an NP-hard problem and is highly intractable to tackle. The NP-hardness primarily stems from the two non-convex constraints \eqref{sec_rate} and \eqref{shift}.
		First, the secrecy rate constraint \eqref{sec_rate} has a difference-of-logarithms structure, and it is generally dealt with by solving a difference-of-convex (DC) programming problem which is NP-hard \cite{lipp2016variations}. Second, the UMC \eqref{shift} restricts the phase shift of each RIS element to the unit circle of the complex plane, forming a non-convex manifold optimization problem which is also NP-hard \cite{Feng2021Optimization}.
		Moreover, the complicated coupling between matrices $\mathbf{W}$ and $\mathbf{\Theta}$ in constraint \eqref{sec_rate} further hinders problem-solving.}

{  In order to tackle Problem \eqref{optimal problem}, we propose a BCD algorithm to divide the optimization variables into three blocks, namely, the number of local computation bits and transmit power $\{\mathbf{l}, \mathbf{p}\}$ at the users, the MUD matrix $\mathbf{W}$ at the AP, and the phase shift matrix $\mathbf{\Theta}$ at the RIS. Based on this, we decompose the original problem into three subproblems which are solved iteratively, and in each subproblem we design one of the above three blocks while fixing the other two.} In the following subsections, we will detail the optimization procedure.		
	\subsection{Joint Optimization of $\{\mathbf{l}, \mathbf{p}\}$ while Fixing $ \mathbf{W} $ and $ \mathbf{\Theta} $}
	Given an MUD matrix $ \mathbf{W} $ and an RIS phase shift matrix $\mathbf{\Theta} $, Problem \eqref{optimal problem} can be simplified to
	\begin{equation}
	\label{compute}
	\underset{\mathbf{l},\mathbf{p}}{\mathop{\mathrm{min}}}~E,\quad\mathrm{s.t.}~\eqref{sec_rate}-\eqref{bits}.
	\end{equation}
	
	Problem (\ref{compute}) is still non-convex due to the non-convex constraint \eqref{sec_rate}. An SCA algorithm is developed to tackle the above problem. To proceed, we first define
	\begin{equation}\label{def}
		\sigma_k = \frac{\sigma_a^2}{\left|\mathbf{w}_k^{\rm H}\mathbf{h}_k\right|^2},~
		\eta_k =\frac{\sigma_e^2}{\left\|\mathbf{g}_{k}\right\|^2} , ~\delta_{k,i} = \frac{\left|\mathbf{w}_k^{\rm H}\mathbf{h}_i\right|^2}{\left|\mathbf{w}_k^{\rm H}\mathbf{h}_k\right|^2},
	\end{equation}
	then the secrecy rate in \eqref{ach_rate} can be reformulated as
	\begin{equation}
		\label{SINR}
		R_{s,k} = \left[\mathrm{log}_2\frac{\eta_k\left(p_k+\sigma_k+\Gamma_k\right)}{\left(\sigma_k+\Gamma_k\right)\left(\eta_k+p_k\right)}\right]^+ = c_k^+(\mathbf{p})-c_k^-(\mathbf{p}),
	\end{equation}
	where
	\begin{equation}
		\Gamma_{k}=\left\{\begin{array}{ll}
			\sum_{i=k+1}^{K} \delta_{k, i} p_{i}, & k \neq K \\
			0, & k=K
		\end{array}\right.
	\end{equation}
	denotes the relative inter-user interference, and we have
	\begin{align}
		c_k^+(\mathbf{p})&=\mathrm{log}_2\left(\eta_k\left(p_k+\sigma_k+\Gamma_k\right)\right),\\
		c_k^-(\mathbf{p})&=\mathrm{log}_2\left(\left(\sigma_k+\Gamma_k\right)\left(\eta_k+p_k\right)\right),\\
		c_k^+(\mathbf{p})&\ge c_k^-(\mathbf{p}).
	\end{align}
	
	Accordingly, constraint \eqref{sec_rate} can be reformulated as
	\begin{equation}		\label{Conc1}
		c_k^+(\mathbf{p})-c_k^-(\mathbf{p}) \ge r_k,~
		c_k^+(\mathbf{p})\ge c_k^-(\mathbf{p}),
	\end{equation}
	where $ r_k \triangleq \left({L_k-l_k}\right)/(BT) $ denotes the target secrecy rate.
	
	Since $ c_k^+(\mathbf{p}) $ and $ c_k^-(\mathbf{p}) $ are concave functions of $ \mathbf{p} $, constraint (\ref{Conc1}), which appears in the form of a subtraction of concave functions, is generally non-convex. In order to tackle this constraint, we use the first-order Taylor expansion to approximate $c_k^-(\mathbf{p})$ as  $F(c_k^-(\mathbf{p}^o)) = c_{k}^{-}\left(\mathbf{p}^o\right)+\nabla_{p} c_{k}^{-}\left(\mathbf{p}^o\right)  \left(\mathbf{p}-\mathbf{p}^o\right) $ for any given $ \mathbf{p}^o $. Then, we can obtain the following convex problem, which can be solved using convex optimization software tools such as CVX. 
	\begin{subequations}
		\label{opt1}
		\begin{align}
			\underset{\mathbf{l},\mathbf{p}}{\mathop{\mathrm{min}}} ~&E,\\
			\mathrm{s.t.} 	~& c_{k}^{+}(\mathbf{p})-F(c_k^-(\mathbf{p}^o)) \ge r_k,~\forall k\in \mathcal{K},\\
			\label{Concave}
			&\eqref{power},\eqref{bits}.
		\end{align}
	\end{subequations}
	
	\subsection{Optimization of $ \mathbf{W} $ while Fixing $\{\mathbf{l}, \mathbf{p}\}$ and $ \mathbf{\Theta} $} 
	Given local computing bits vector $ \mathbf{l} $, transmit power vector $ \mathbf{p} $, and phase shift matrix $ \mathbf{\Theta} $,	Problem \eqref{optimal problem} becomes a feasibility checking problem, which can be reformulated as 
	\begin{equation}
		\label{w}
		\mathrm{find} ~	 \mathbf{W},\quad  \mathrm{s.t.} ~\eqref{sec_rate}.
	\end{equation}

To address the non-convex constraint \eqref{sec_rate}, a natural approach is to reformulate Problem (\ref{w}) as an SDP problem using the matrix lifting technique \cite{Alavi2018Beamforming}. First lift \( \mathbf{W}_k \triangleq \mathbf{w}_k \mathbf{w}_k^{\rm H}\) with \( \mathrm{rank}(\mathbf{W}_k) = 1, k \in \mathcal{K} \), and then we apply SDR to relax the non-convex rank-one constraint. By doing this, Problem (\ref{w}) can be decoupled into \( K \) sub-problems as follows:
	\begin{subequations}
		\label{w1}
		\begin{align}
			\mathrm{find} ~ &\mathbf{W}_k,\\
			\mathrm{s.t.} ~	
			&\dfrac{p_k \mathrm{Tr}(\mathbf{W}_k \mathbf{H}_k) }{\sum_{i=k+1}^K p_i \mathrm{Tr}(\mathbf{W}_k\mathbf{H}_i )  +\sigma_a^2  }  \ge 2^{r_k}\left(1+\frac{p_k}{\eta_k} \right) -1,\\
			&\mathbf{W}_k \succeq 0, \mathrm{Tr}\left(\mathbf{W}_k\right) = 1,
		\end{align}
	\end{subequations}
	where $ \mathbf{H}_k \triangleq \mathbf{h}_k\mathbf{h}_k^{\rm H} $, $ k\in \mathcal{K} $, and $\eta_k$ is defined in \eqref{def}. Note that for a fixed phase shift matrix $ \mathbf{\Theta} $, the equivalent channel gains $ \mathbf{h}_k $ and $ \mathbf{g}_k $, as well as $\mathbf{H}_k$ and $\eta_k$, are also fixed. 
	
	It can be observed that Problem (\ref{w1}) is a convex optimization problem and can be solved using CVX. Generally, the relaxed Problem (\ref{w1}) may not lead to a rank-one solution, which implies that the solution to Problem \eqref{w1} only serves as a lower bound of (\ref{w}). Thus, we use the Gaussian randomization technique to construct a rank-one solution from the obtained higher-rank solution to Problem (\ref{w1}) \cite{Wu2019Intelligent}.
	{ \subsection{Optimization of $\mathbf{\Theta} $ while Fixing $\{\mathbf{l}, \mathbf{p}\}$ and $ \mathbf{W} $ } 
	For fixed $\{\mathbf{l}, \mathbf{p}\}$ and $\mathbf{W}$, the phase shift matrix $ \mathbf{\Theta} $ should meet both the secrecy rate constraint \eqref{sec_rate} and UMC \eqref{shift}, yielding the following problem 
		\begin{equation}
		\label{thata}
		\mathrm{find} ~ \mathbf{\Theta},\quad\mathrm{s.t.} ~	\eqref{sec_rate},  \eqref{shift}.	
	\end{equation} 
	
	In order to address the above problem efficiently, we adopt quadratic form reformulation \cite{Wu2019Intelligent} to modify constraint $\eqref{sec_rate}$ by expressing secrecy rate $R_{s,k}$ more explicitly. Specifically, we introduce $ \mathbf{v}_{i,k} \triangleq \mathbf{w}_i^{\rm H}\mathbf{H} \mathrm{diag}(\mathbf{h}_{r,k}) $ and $ q_{i,k} \triangleq   \mathbf{w}_i^{\rm H}\mathbf{h}_{a,k} $, and meanwhile we define $ \mathbf{\Theta} = \mathrm{diag}\left(\mathbf{e}\right) $ with $ \mathbf{e} \triangleq \left[e_1, \cdots, e_M\right]^{\rm T}$ and $e_m\triangleq e^{j\theta_m} $. In this way, the term $\left|\mathbf{w}_i^{\rm H}\mathbf{h}_{k}\right|^2= \left|\mathbf{w}_i^{\rm H}( \mathbf{H} \mathbf{\Theta}\mathbf{h}_{r,k} + \mathbf{h}_{a,k})\right|^2 $ associated with AP's SINR $\gamma_k$ in \eqref{user_SINR} can be rewritten as $ \left|\mathbf{v}_{i,k} \mathbf{e}  + q_{i,k}  \right|^2 $. By augmenting \( \mathbf{e} \) with an additional dimension as \( \overline{\mathbf{e}} \triangleq  [\mathbf{e}; 1] \), the term $ \left|\mathbf{v}_{i,k} \mathbf{e}  + q_{i,k} \right|^2 $ can be further expressed as $	\overline{\mathbf{e}}^{\rm H} \mathbf{V}_{i,k} \overline{\mathbf{e}} + \left|q_{i,k}\right|^2,$
where we have
	\begin{equation}
		\mathbf{V}_{i,k} \triangleq \begin{bmatrix}
			\mathbf{v}_{i,k}^{\rm H} \mathbf{v}_{i,k} & \mathbf{v}_{i,k}^{\rm H} q_{i,k} \\
			\mathbf{v}_{i,k} q_{i,k}^{\dagger} & 0
		\end{bmatrix}, \quad\forall k\in \mathcal{K}.
	\end{equation}
	
	Similarly, for the term $ \|\mathbf{g}_k\|^2= \left\|\mathbf{G}\mathbf{\Theta}\mathbf{h}_{r,k} + \mathbf{h}_{e,k}\right\|^2$ associated with Eve's SINR $\gamma_{e,k}$ in \eqref{eve_SINR}, we introduce two symbols
	\begin{equation}
		\mathbf{J}_{e,k} \triangleq \begin{bmatrix}
			\mathrm{diag}(\mathbf{h}_{r,k}^{\rm H})\mathbf{G}^{\rm H} \\
			\mathbf{h}_{e,k}^{\rm H}
		\end{bmatrix}, ~\mathbf{S}_{e,k} \triangleq \mathbf{J}_{e,k}\mathbf{J}_{e,k}^{\rm H},\quad\forall k\in \mathcal{K},
	\end{equation}
	and the term $ \|\mathbf{g}_k\|^2$ can be rewritten as \( \overline{\mathbf{e}}^{\rm H} \mathbf{S}_{e,k} \overline{\mathbf{e}} \) with $ k \in \mathcal{K} $.
	 
	To simplify the optimization procedure, we define $ \mathbf{E} \triangleq  \overline{\mathbf{e}}\overline{\mathbf{e}}^{\rm H} $ such that 
$	\overline{\mathbf{e}}^{\rm H} \mathbf{V}_{i,k} \overline{\mathbf{e}} = \mathrm{Tr}(\mathbf{V}_{i,k} \mathbf{E})$ and $\overline{\mathbf{e}}^{\rm H} \mathbf{S}_k \overline{\mathbf{e}} = \mathrm{Tr}(\mathbf{S}_k \mathbf{E})$ with \( \mathbf{E} \succeq \mathbf{0} \) and \( \mathrm{rank}(\mathbf{E}) = 1 \). We further define $	 \overline{\mathbf{V}}_{i,k} \triangleq \mathrm{Tr}(\mathbf{V}_{i,k} \mathbf{E}) + |q_{i,k}|^2$ and $ \overline{\mathbf{S}}_{e,k} \triangleq \mathrm{Tr}(\mathbf{S}_{e,k} \mathbf{E})$, then Problem \eqref{thata} can be recast as below upon using SDR to relax the rank-one constraint $ \mathrm{rank}(\mathbf{E}) = 1$.
	\begin{subequations}
		\label{the}
		\begin{align}
		 \mathrm{find} &~\mathbf{E}, \\
			\mathrm{s.t.} 			\label{rate}
			&~\mathrm{log}_2 \left( 1 + \dfrac{p_k \overline{\mathbf{V}}_{k,k}}{\sum_{i=k+1}^K p_i \overline{\mathbf{V}}_{i,k} + \sigma_a^2} \right) - \mathrm{log}_2 \left( 1 + \dfrac{p_k \overline{\mathbf{S}}_{e,k}}{\sigma_e^2} \right) \nonumber \\
			&~ \quad\quad\quad\quad\quad\quad\quad\quad \ge r_k, ~\forall k\in \mathcal{K},\\
			\label{zhi}
		&~\mathbf{E}\left[m,m\right] = 1,~\forall  m\in \left\{1, \cdots, M+1\right\},~\mathbf{E} \succeq \mathbf{0},
		\end{align}
	\end{subequations}
	where $\mathbf{E}\left[m,m\right]$ is the $m$-th diagonal element of $\mathbf{E}$.
}
	 
	The above problem remains non-convex due to the non-convexity of constraint \eqref{rate}. To address this issue, the following lemma is utilized \cite{Christensen2008Weighted}.
	\begin{lemma}
		\label{lemma1}
		For a function \( y(\mu) = -\mu x + \mathrm{ln} \mu + 1 \), \( \forall x > 0 \), 
		\begin{equation}
			-\mathrm{ln}x = \underset{\mu>0}{\mathop{\mathrm{max}}}~y(\mu),
		\end{equation}
	\end{lemma}
	where the optimal solution is $ \mu = 1/x $.
	
	Based on Lemma 1 and letting $ x = \sum_{i=k+1}^K p_i \overline{\mathbf{V}}_{i,k}  +\sigma_a^2 $ and $ \mu = \mu_k $, the first term of constraint \eqref{rate} can be given as
	\begin{equation}
		\label{phi_k}
		\begin{aligned}
			&\mathrm{log}_2\left( \sum_{i=k}^K p_i \overline{\mathbf{V}}_{i,k} + \sigma_a^2 \right) - \mathrm{log}_2\left( \sum_{i=k+1}^K p_i \overline{\mathbf{V}}_{i,k} + \sigma_a^2 \right) \\
			&= \underset{\mu_k > 0}{\mathop{\mathrm{max}}} \, \phi_k(\mathbf{E}, \mu_k),
		\end{aligned}
	\end{equation}
	where 
	\begin{align}
		\phi_k(\mathbf{E}, \mu_k) = \frac{1}{\ln 2} \Bigg[
		\ln & \left( \sum_{i=k}^K p_i \overline{\mathbf{V}}_{i,k} + \sigma_a^2 \right) - \mu_k \sum_{i=k+1}^K p_i \overline{\mathbf{V}}_{i,k} \nonumber \\
		&- \mu_k \sigma_a^2 + \ln\mu_k + 1 
		\Bigg].
	\end{align}
	
	Similarly, let $ x = p_k \overline{\mathbf{S}}_{e,k} + \sigma_e^2  $ and $\mu = \mu_{e,k}$, the second term of constraint \eqref{rate} can be written as
	\begin{equation}
	 \frac{\mathrm{ln}\sigma_e^2-\mathrm{ln}\left( p_k \overline{\mathbf{S}}_{e,k} + \sigma_e^2 \right) }{\mathrm{ln}2} = \underset{\mu_{e,k}>0}{\mathop{\mathrm{max}}}~\phi_{e,k}(\mathbf{E},\mu_{e,k}),
	\end{equation}
	where 
		\begin{equation}
		\phi_{e,k}(\mathbf{E}, \mu_{e,k}) = \frac{1- \mu_{e,k} \left( p_k \overline{\mathbf{S}}_{e,k} + \sigma_e^2 \right)			+ \ln \mu_{e,k}  + \ln \sigma_e^2}{\ln 2}.
		\end{equation}

	According to Lemma 1, for a fixed $ \mathbf{E} $, the optimal values of $ \mu_k $ and $ \mu_{e,k} $ can be achieved when $ \mu_k^{*} = 1/\left(  \sum_{i=k+1}^K p_i \overline{\mathbf{V}}_{i,k}  +\sigma_a^2 \right)  $ and $ \mu_{e,k}^{*} = 1/\left( p_k \overline{\mathbf{S}}_{e,k} + \sigma_e^2\right) $. Thus, Problem \eqref{the} can be converted into:
	\begin{subequations}
		\label{fin}
		\begin{align}
			\mathrm{find} ~&\mathbf{E},\\ 
	\mathrm{s.t.}~	& 
			\phi_k(\mathbf{E},\mu_k^*) + \phi_{e,k}(\mathbf{E},\mu_{e,k}^*) \ge r_k,\quad\forall k\in \mathcal{K},\label{offloading_rate}\\
			& \eqref{zhi},\label{re_zhi}
		\end{align}
	\end{subequations}
	where $\mu_k^*$ and $\mu_{e,k}^*$ are calculated based on the optimal $\bE$ obtained from the last iteration.
	  
	In order to solve Problem \eqref{fin} more efficiently, it would be better to introduce an explicit objective function. The rationale is that for the energy consumption minimization Problem \eqref{optimal problem}, the secrecy rate constraint \eqref{sec_rate} for each user should be active at the optimal solution. As such, optimizing the RIS phase shifts is equivalent to enforcing the achievable secrecy rate to exceed the targeted offloading rate as much as possible as indicated in \eqref{offloading_rate}. { To this end, we introduce a slack variable $\nu_k$ to quantify the rate residual of user $k$, i.e., the difference between the secrecy rate and offloading rate,
	\begin{equation}
		 \label{residual_rate}
		\nu_k = \phi_k(\mathbf{E},\mu_k^*) + \phi_{e,k}(\mathbf{E},\mu_{e,k}^*) - r_k,\quad\forall k\in \mathcal{K}.
	\end{equation}
	
With \eqref{residual_rate}, Problem \eqref{fin} can be transformed as 
	\begin{equation}
		\label{fin1}
			\underset{\mathbf{E}}{\mathop{\mathrm{max}}}
\underset{k\in \mathcal{K}}{\sum}\nu_k,\quad\mathrm{s.t.}~~ \eqref{offloading_rate},\eqref{re_zhi}.
	\end{equation}
}
	So far, Problem \eqref{fin1} is convex and can be solved by employing CVX. After obtaining the optimal $ \mathbf{E} $, the Gaussian randomization technique is applied to construct a rank-one solution of $ \overline{\mathbf{e}} $, denoted as $ \tilde{\mathbf{e}} $. Finally, following the approach in \cite{Li2022Intelligent}, the phase shifts can be determined as
	\begin{equation}
		\theta_m = \angle\left(\frac{\tilde{\mathbf{e}}[m]}{\tilde{\mathbf{e}}[M+1]}\right),
	\end{equation}
	where $ \angle(x) $ denotes the phase of $ x $, and $\tilde{\mathbf{e}}[m]$ denotes the $m$-th element of $ \tilde{\mathbf{e}} $.

	{ Note that BCD algorithms inherently guarantee convergence under the conditions of subproblem convexity and monotonic improvement \cite{boyd2004convex}. In our proposed BCD algorithm, each iterated subproblem is transformed to be convex after applying SCA and SDR and meanwhile yields a lower total energy consumption, thereby guaranteeing the convergence.} The overall algorithm is summarized as Algorithm 1.

	\begin{algorithm}
		\caption{BCD Algorithm for solving Problem \eqref{optimal problem}}
		\label{algorithm1}
		\hspace{0.02in}{\bf Input:} $B, C_k, T, \varsigma_k, K, \mathbf{h}_{a,k}, \mathbf{h}_{e,k}, \mathbf{h}_{r,k}, \mathbf{G}, \mathbf{H}, \varepsilon$.\\
		\hspace{0.02in}{\bf Output:} $\mathbf{l}^*, \mathbf{p}^*, \mathbf{W}^*, \mathbf{\Theta}^*$. 
		\begin{algorithmic}[1]
			\State Initialize $\mathbf{l}^{(0)}$, $\mathbf{p}^{(0)}$, $\mathbf{W}^{(0)}$, $\mathbf{\Theta}^{(0)}$, calculate $E^{(0)}$, and set iteration number $i = 0$.
			\Repeat
			\State Solve Problem \eqref{opt1} using CVX toolbox with given $\mathbf{W}^{(i)}$ and $\mathbf{\Theta}^{(i)}$, and obtain the optimal $\mathbf{l}^{(i+1)}$ and $\mathbf{p}^{(i+1)}$.
			\State Solve Problem \eqref{w1} with given $\mathbf{l}^{(i+1)}$, $\mathbf{p}^{(i+1)}$, and $\mathbf{\Theta}^{(i)}$, and denote the solution after Gaussian randomization as $\mathbf{W}^{(i+1)}$.
			\State Solve Problem \eqref{fin1} with given $\mathbf{l}^{(i+1)}$, $\mathbf{p}^{(i+1)}$, and $\mathbf{W}^{(i+1)}$, and denote the solution after Gaussian randomization as $\mathbf{\Theta}^{(i+1)}$.
			\State Calculate $E^{(i+1)}$ according to \eqref{energy_equation} and update $i = i+1$.
			\Until{$\frac{\left| E^{(i)} - E^{(i-1)} \right|}{\left| E^{(i-1)} \right|} \le \varepsilon$}.
		\end{algorithmic}
	\end{algorithm}
	
{ The overall computational complexity of Algorithm \ref{algorithm1} arises from solving the three subproblems \eqref{opt1}, \eqref{w1}, and \eqref{fin1}. 
In order to analyze the complexity of each subproblem, we first give the computational complexity of solving a convex optimization problem involving $I$ second-order cone (SOC) constraints and $j$ linear matrix inequality (LMI) constraints using an interior-point method \cite{wang2014outage}, which is given below,
		\begin{equation} 
		\label{complexity}
		\mathcal{O}\left( \ln \frac{1}{\varepsilon}
		\left[
		n^3 +\underbrace{n \sum_{j=1}^{J} \tau_j^2\left(\tau_j+n\right)}_{\text{due to LMI}}  + \underbrace{n \sum_{i=1}^{I} \varphi_i^2} _{\text{due to SOC}}
		\right] \sqrt{\sum_{j=1}^{J} \tau_j + 2I}
		\right)
		\end{equation}
where \(n\) denotes the number of optimization variables, \(\tau_j \times \tau_j\) is the size of the symmetric matrix involved in the \(j\)-th LMI, and \(\varphi_i\) is dimension of the \(i\)-th SOC constraint.

Based on \eqref{complexity}, we have \(n=2K\), \(J=3K\), \(I=0\) for Problem \eqref{opt1}, \(n=2N_a^2\), \(J=2\), \(I=0\) for Problem~\eqref{w1}, and \(n=M^2\), \(J=K\), \(I=0\) for Problem~\eqref{fin}. By substituting these parameter configurations into \eqref{complexity} and omitting the logarithmic term \(\ln \frac{1}{\varepsilon}\), we can calculate the per-iteration complexities for solving subproblems \eqref{opt1}, \eqref{w1}, and \eqref{fin1}, which are $o_{\mathbf{lp}} = \mathcal{O}\left( K^{3.5}\right)$, 
$o_{\mathbf{W}}=\mathcal{O}\left( N_a^{6.5}K\right)$, and $o_{\mathbf{E}}=\mathcal{O}\left( K^{1.5}M^{6.5}\right)$, respectively.
Therefore, the overall per-iteration complexity of solving Problem \eqref{optimal problem} can be given by
\begin{equation}
	o_{\text{tot}} = o_{\mathbf{lp}}+ o_{\mathbf{W}} + o_{\mathbf{E}}.
\end{equation}}
	
	\section{Robust Secure Offloading with Imperfect CSI}
		The performance achieved under the perfect CSI case can serve as an upper bound for the proposed secure communication design. However, since Eve normally tries to hide her existence from the AP, it is difficult to obtain the perfect CSI between Eve and the other nodes. In this section, the total energy consumption minimization problem is extended to the case with imperfect estimation of Eve's channels $ \mathbf{G}_k $ and $ \mathbf{h}_{e,k} $. Similar to the previous section, the local computing bits vector, the transmit power vector, the MUD matrix, and the phase shift matrix are jointly optimized to maximize energy efficiency.

	\subsection{Problem Formulation}
	We assume that the actual channel lies within a bounded uncertainty region around a nominal (estimated) channel. This approach guarantees system performance under the worst possible realization of the channel within this region. For the reflecting link, we denote $ \mathbf{G}_k = \mathbf{G}\mathrm{diag}\left( \mathbf{h}_{r,k} \right) $ as the cascaded channel from the user $ k $ to Eve via the RIS, and $ \mathbf{e}=\left[e_{1}, \ldots, e_{M}\right]^{\mathrm{T}}  $ as the vector containing the diagonal elements of the matrix $ \mathbf{\Theta} $. The channel estimation error is modeled as the bounded CSI error which is given by
	\begin{align}
		\mathbf{h}_{e,k}&=\overline{\mathbf{h}}_{e,k} + \triangle \mathbf{h}_{e,k},\\
		\mathbf{G}_k &= \overline{\mathbf{G}}_k +  \triangle \mathbf{G}_k,\\
		{{\Omega }_{e,k}}&=\left\{ {{\left\| {\triangle \mathbf{h}_{e,k}} \right\|}_{2}}\le {{\epsilon }_{e,k}},{{\left\| \triangle \mathbf{G}_k \right\|}_{F}}\le {{\epsilon }_{g,k}} \right\},
	\end{align}
	where $ \overline{\mathbf{h}}_{e,k} $ and $ \overline{\mathbf{G}}_k $ are the estimated channels of $\mathbf{h}_{e,k} $ and $ \mathbf{G}_k $, respectively, $ {{\epsilon }_{e,k}} > 0 $ and $ {{\epsilon }_{g,k}} > 0 $ denote the sizes of the uncertainty regions of the channel estimation errors $\triangle \mathbf{h}_{e,k} $ and $ \triangle \mathbf{G}_k $, respectively.\footnote{{The bounded CSI error model has been widely adopted for robust system design due to mathematical tractability and suitability for the scenario with limited statistical knowledge, such as RIS-aided secure MEC networks against passive eavesdroppers \cite{li2022secure}. In practice, the error bounds \(\epsilon_{e,k}\) and \(\epsilon_{g,k}\) can be obtained via channel estimation or quantization processes. For example, the estimation error can be approximated by  \(\epsilon_{e,k}\approx\sqrt{\epsilon_{\rm MSE}N_e }\) for least-squares (LS) channel estimation, where \(\epsilon_{\rm MSE} \) is the mean squared error derived from the pilot signal-to-noise ratio (SNR); the quantization error can be bounded as \(\epsilon_{e,k} \leq 2^{-B/(N_e-1)}\sqrt{N_e}\) for Grassmannian codebook-based quantization, where \(B\) feedback bits are used for channel vector quantization.}}

	{ In order to develop a robust secure offloading scheme with imperfect Eve's CSI, we consider a worst-case scenario with the worst estimate of Eve's channels thus resulting in a minimal secrecy rate. Recalling \eqref{ach_rate}, the worst-case achievable secrecy rate of user $k$, $\forall k\in \mathcal{K}$, can be expressed as
	\begin{equation}
		R_{s,k} = \left[ \mathrm{log}_2\left(1+\gamma_k\right)-  \underset{\Omega_{e,k}}{\mathrm{max}}\:\mathrm{log}_2\left(1+\gamma_{e,k}\right)\right]^+,
	\end{equation}
	where $\gamma_k$ and $\gamma_{e,k}$ denote AP's and Eve's SINRs for decoding the signal from user $ k $, as defined in 
	\eqref{user_SINR} and \eqref{eve_SINR}, respectively.}
	
	By considering the imperfect CSI model, the total energy consumption minimization problem can be formulated as 
    	\begin{subequations}
	{ 	\label{im_CSI}
		\begin{align}
			\underset{\mathbf{l},\mathbf{p},\mathbf{W},\mathbf{e}}{\mathrm{min}}
			~& E = \underset{k\in \mathcal{K}}{\sum}\left(\frac{{{\varsigma }_k}{C_k^3}{l_k^3}}{T^2}+{p_kT}\right),\\
			\mathrm{s.t.} ~
			\label{secrecy rate}
			& R_{s,k} \ge r_k, \quad \forall k \in \mathcal{K}, \\
			\label{im_power}
			& 0 \le p_k \le P_k, \quad \forall k \in \mathcal{K}, \\
			\label{con_power}
			& 0 \le l_k \le L_k, \quad \forall k \in \mathcal{K}, \\
			\label{phase}
			& \left| e_m \right| = 1, \quad \forall m \in \mathcal{M},
		\end{align}}
	\end{subequations}
	
	Problem \eqref{im_CSI} is a challenging NP-hard problem due to the non-convex secrecy rate constraint \eqref{secrecy rate} and UMC \eqref{phase}, as well as the complicated coupling between $\mathbf{W}$ and $\mathbf{\Theta}$ in constraint \eqref{secrecy rate}.
	Inspired by Problem \eqref{optimal problem}, we propose a BCD algorithm to solve this problem, which is detailed as below.

	\subsection{Joint Optimization of $\{\mathbf{l}, \mathbf{p}\}$ while Fixing $ \mathbf{W} $ and $ \mathbf{e}$}
		In this subsection, we solve Problem \eqref{im_CSI} by designing the optimal local computing bits vector $ \mathbf{l} $ and transmit power vector $ \mathbf{p} $ for given $ \mathbf{W} $ and $ \mathbf{e}$. Note that the major difficulty lies in the non-convex secrecy rate constraint \eqref{secrecy rate}.
	
	To start with, constraint \eqref{secrecy rate} can be formulated as follows by introducing an auxiliary variable $ a_k $ as the upper bound of Eve's received SNR of user $k$ for $\forall k \in \mathcal{K}$,
	\begin{subequations}
		\begin{align}
			\label{im_a1}
			\mathrm{log}_2\left(1 + \dfrac{p_k \left| \mathbf{w}^{\rm H}_k \mathbf{h}_k \right|^2 }{\sum_{j=k+1}^K p_j \left| \mathbf{w}^{\rm H}_k \mathbf{h}_j \right|^2 + \sigma_a^2 }\right) 
			&- \mathrm{log}_2(1 + a_k) \nonumber \\
			& \ge r_k, \\
			\label{im_a2}
			p_k \left\| \mathbf{h}_{e,k} + \mathbf{G}_k \mathbf{e} \right\|^2
			&\le a_k \sigma_e^2.
		\end{align}
	\end{subequations}

	Based on Lemma 1, let \( x = \sum_{j=k+1}^K p_j \left| \mathbf{w}^{\rm H}_k \mathbf{h}_j \right|^2 + \sigma_a^2 \) and \( \mu = \mu_{a,k} \). The first term of constraint \eqref{im_a1} can be written as
	\begin{align}
	\label{im_b1}
	&\mathrm{log}_2\left( \sum_{j=k}^K p_j \left| \mathbf{w}^{\rm H}_k \mathbf{h}_j \right|^2 + \sigma_a^2 \right)  
	-\mathrm{log}_2\left( \sum_{j=k+1}^K p_j \left| \mathbf{w}^{\rm H}_k \mathbf{h}_j \right|^2 + \sigma_a^2 \right)\nonumber\\
	&	
	= \underset{\mu_{a,k} > 0}{\mathop{\mathrm{max}}} \: \psi_{a,k} (p_k, \mu_{a,k}).
	\end{align}
	where 
	\begin{align}
		\psi_{a,k}&(p_k, \mu_{a,k}) = \frac{1}{\ln 2} \Bigg[ 
		\ln \left( \sum_{j=k}^K p_j \left| \mathbf{w}^{\rm H}_k \mathbf{h}_j \right|^2 + \sigma_a^2 \right) -\nonumber \\
		& \mu_{a,k} \left( \sum_{j=k+1}^K p_j \left| \mathbf{w}^{\rm H}_k \mathbf{h}_j \right|^2 + \sigma_a^2 \right) + \ln\mu_{a,k} + 1 \Bigg].
	\end{align}

	Similarly, the second part of constraint \eqref{im_a1} can be transformed to
	\begin{equation}
		-\mathrm{log}_2( 1+a_k)= \underset{\mu_{e,k}>0}{\mathop{\mathrm{max}}} \: \psi_{e,k} (a_k,\mu_{e,k}),
	\end{equation}
	where $\psi_{e,k} (a_k,\mu_{e,k})=\left[ 1- \mu_{e,k}\left( 1+a_k\right)+\mathrm{ln}\mu_{e,k} \right]/\ln2 $.
	
	{ As for constraint \eqref{im_a2}, we introduce an auxiliary variable $ \beta_k $ to represent an upper bound of the channel gain $\left\| \mathbf{h}_{e,k} + \mathbf{G}_k\mathbf{e}\right\|^2$, then  \eqref{im_a2} can be transformed into }
	\begin{subequations}
		\begin{align}
			\label{t1}
			\left\| \mathbf{h}_{e,k} + \mathbf{G}_k\mathbf{e}\right\|^2 &\le \beta_k,\\
			\label{t2}
			p_k\beta_{k} &\le {a_k}\sigma_e^2.
		\end{align}
	\end{subequations}
	
	In order to deal with the uncertainty in $ \left\lbrace \triangle \mathbf{h}_{e,k}, \triangle \mathbf{G}_k \right\rbrace  $ of constraint \eqref{t1}, we first adopt Schur's complement lemma to equivalently recast constraint \eqref{t1} into matrix inequalities,
	\begin{equation}
		\label{schur}
		\left[\begin{array}{cc}
			\beta_k &\mathbf{g}_k^{\rm H} \\
			\mathbf{g}_k &\mathbf{I}
		\end{array}\right] \succeq \mathbf{0}.
	\end{equation}
	
	Substituting $ \mathbf{h}_{e,k}=\overline{\mathbf{h}}_{e,k} + \triangle \mathbf{h}_{e,k}$ and $\mathbf{G}_k = \overline{\mathbf{G}}_k + \triangle \mathbf{G}_k $ into constraint \eqref{schur} leads to:
	\begin{equation}
		\label{deco}
		\begin{aligned}
			\mathbf{0} \preceq & \left[ \begin{array}{cc}
				\beta_k & \widetilde{\mathbf{t}}_k^{\rm H} \\
				\widetilde{\mathbf{t}}_k & \mathbf{I}_{N_e}
			\end{array} \right] +
			\left[ \begin{array}{cc}
				0 & \triangle \mathbf{h}_{e,k}^{\rm H} + \mathbf{e}^{\rm H} \triangle \mathbf{G}_k^{\rm H} \\
				\triangle \mathbf{h}_{e,k} + \triangle \mathbf{G}_k \mathbf{e} & \mathbf{0}_{N_e \times N_e}
			\end{array} \right] \\
			= & 
			\left[\begin{array}{c}
				\mathbf{0}_{1 \times N_e} \\
				\mathbf{I}_{N_e}
			\end{array} \right] \left[ \triangle \mathbf{h}_{e,k} \quad \mathbf{0}_{N_e \times N_e} \right] 
			+
			\left[ \begin{array}{c}
				\triangle \mathbf{h}_{e,k}^{\rm H} \\
				\mathbf{0}_{N_e \times 1}
			\end{array} \right] \left[ \mathbf{0}_{1 \times N_e} \quad \mathbf{I}_{N_e} \right] \\
			& +
			\left[ \begin{array}{cc}
				\beta_k & \widetilde{\mathbf{t}}_k^{\rm H} \\
				\widetilde{\mathbf{t}}_k & \mathbf{I}_{N_e}
			\end{array} \right]
			+ 
			\left[ \begin{array}{c}
				\mathbf{0}_{1 \times M} \\
				\mathbf{I}_{N_e}
			\end{array} \right] \triangle \mathbf{G}_k 
			\left[ \mathbf{e} \quad \mathbf{0}_{M \times N_e} \right] \\
			& +
			\left[ \begin{array}{c}
				\mathbf{e}^{\rm H} \\
				\mathbf{0}_{N_e \times M}
			\end{array} \right] 
			\triangle \mathbf{G}_k^{\rm H}
			\left[ \mathbf{0}_{N_e \times 1} \quad \mathbf{I}_{N_e} \right],
		\end{aligned}
	\end{equation}
	where $ \widetilde{\mathbf{t}}_{k}\triangleq \overline{\mathbf{h}}_{e,k} + \overline{\mathbf{G}}_k\mathbf{e} $, $ \mathbf{I}_{l} $ and $\mathbf{0}_{m,n}$ denote the $ l\times l $ identity matrix and $m\times n$ zero matrix, respectively. To simplify the matrix inequality constraint in \eqref{deco}, the following general sign-definiteness principle is applied \cite{Gharavol2012Sign}.
	\begin{lemma}[General Sign-Definiteness]
		\label{sign}
		For matrices $ \mathbf{M}=\mathbf{M}^{\rm H} $ and $ \left\{\mathbf{Y}_{i}, \mathbf{Z}_{i}\right\}_{i=1}^{P} $, the linear matrix inequality (LMI) 
		\begin{equation}
			\mathbf{M} \succeq \sum_{i=1}^{P}\left(\mathbf{Y}_{i}^{\mathrm{H}} \mathbf{X}_{i} \mathbf{Z}_{i}+\mathbf{Z}_{i}^{\mathrm{H}} \mathbf{X}_{i}^{\mathrm{H}} \mathbf{Y}_{i}\right),~\forall i,\mathbf{X}_i: \left\|\mathbf{X}_{i}\right\|_{F} \leq \xi_{i}
		\end{equation}
		holds if and only if there exist real numbers $  \mu_{i} \geq 0$ such that
		\begin{equation}
		\left[\begin{array}{cccc}
		\mathbf{M}-\sum_{i=1}^{P} \mu_{i} \mathbf{Z}_{i}^{\mathrm{H}} \mathbf{Z}_{i} & -\xi_{1} \mathbf{Y}_{1}^{\mathrm{H}} & \cdots & -\xi_{P} \mathbf{Y}_{P}^{\mathrm{H}} \\
		-\xi_{1} \mathbf{Y}_{1} & \mu_{1} \mathbf{I}_r & \cdots & \mathbf{0}_{r \times r} \\
		\vdots & \vdots & \ddots & \vdots \\
		-\xi_{P} \mathbf{Y}_{P} & \mathbf{0}_{r \times r} & \cdots & \mu_{P} \mathbf{I}_r
		\end{array}\right] \succeq \mathbf{0}.
		\end{equation}
	\end{lemma}
	
	By applying Lemma \ref{sign}, we choose the following parameters for each constraint in \eqref{deco} as follows:
	\begin{equation}
	\begin{aligned}
	& \mathbf{M} = \begin{bmatrix}
	\beta_{k} & \widetilde{\mathbf{t}}_{k}^{\mathrm{H}} \\
	\widetilde{\mathbf{t}}_{k} & \mathbf{I}_{N_e}
	\end{bmatrix}, \quad P=2, \\
	& \mathbf{Y}_1 = -\begin{bmatrix} \mathbf{0}_{N_e \times 1} & \mathbf{I}_{N_e} \end{bmatrix}, \quad \mathbf{Z}_1 = \mathbf{I}_{N_e + 1}, \quad \mathbf{X}_1 = \triangle \mathbf{h}_{e,k}, \\
	& \mathbf{Y}_2 = -\begin{bmatrix} \mathbf{0}_{N_e \times 1} & \mathbf{I}_{N_e} \end{bmatrix}, \quad \mathbf{Z}_2 = \begin{bmatrix} \mathbf{e} & \mathbf{0}_{M \times N_e} \end{bmatrix}, \quad \mathbf{X}_2 = \triangle \mathbf{G}_k.
	\end{aligned}
	\end{equation}

		Then, the equivalent LMI of constraint \eqref{t1} is given by
		\begin{equation}
		\label{uncert}
		\begin{bmatrix}
		\mathbf{T}_k & -\xi_{\mathrm{e},k} \mathbf{Y}_1^{\mathrm{H}} & -\xi_{\mathrm{g},k} \mathbf{Y}_2^{\mathrm{H}} \\
		-\xi_{\mathrm{e},k} \mathbf{Y}_1 & \mu_{h,k} \mathbf{I}_{N_e} & \mathbf{0}_{N_e\times N_e} \\
		-\xi_{\mathrm{g},k} \mathbf{Y}_2 & \mathbf{0}_{N_e\times N_e} & \mu_{g,k} \mathbf{I}_{N_e}
		\end{bmatrix} \succeq \mathbf{0},
		\end{equation}
		where $ \bT_k = \mathbf{M}-\mu_{\mathrm{h}, k}\bZ_1^{\rm H}\bZ_1 - \mu_{\mathrm{g}, k}\bZ_2^{\rm H}\bZ_2 $, and $ \boldsymbol{\mu}_{\mathrm{g}}=\left[\mu_{\mathrm{g}, 1}, \ldots, \mu_{\mathrm{g}, K}\right]^{\mathrm{T}} $, $ \boldsymbol{\mu}_{\mathrm{h}}=\left[\mu_{\mathrm{h}, 1}, \ldots, \mu_{\mathrm{h}, K}\right]^{\mathrm{T}} $ are slack variables with $\mu_{\mathrm{h}, k} \ge 0$ and $\mu_{\mathrm{g}, k} \ge 0$, respectively.

	However, constraint \eqref{t2} is still non-convex. We further employ SCA to approximate the non-convex term $p_k \beta_k$ with a convex one by applying the first-order Taylor expansion as the upper bound. As a result, constraint \eqref{t2} can be given as 
	\begin{equation}
		\label{plsca}
		\beta_k^{(i)}p_k^{(i)}+\beta_k^{(i)}\left(p_k-p_k^{(i)}\right)+p_k^{(i)}\left(\beta_k - \beta_k^{(i)}\right) \le {a_k}\sigma_e^2, 
	\end{equation}
	where $p_k^{(i)}$ and $\beta_k^{(i)}$ are the optimal values of $p_k$ and $\beta_k$ from the last iteration, respectively. After that, Problem \eqref{im_CSI} for optimizing $\{\mathbf{l}, \mathbf{p}\}$ can be transformed into
	\begin{subequations}
		\label{im_CSI_lp}
		\begin{align}
			\underset{\mathbf{l},\mathbf{p}, \boldsymbol{\mu}_{\mathrm{g}}, \boldsymbol{\mu}_{\mathrm{h}}, \mathbf{a}, \boldsymbol{\beta}}{\mathop{\mathrm{min}}}
			~    
			&E,\\
			\mathrm{s.t.} ~   
			&\phi_{a,k} (p_k,\mu_{a,k})+ \phi_{e,k} (a_k,\mu_{e,k})\ge r_k,\\
			& \eqref{im_power},\eqref{con_power}, \eqref{uncert}, \eqref{plsca},\\
			\label{mu}
			& \mu_{\mathrm{g},k} \ge 0, ~\mu_{\mathrm{h},k} \ge 0,
		\end{align}
	\end{subequations}
	where $\mathbf{a} \triangleq \left[a_1, \dots, a_K\right]$ and $\boldsymbol{\beta} \triangleq \left[\beta_1,\dots,\beta_K\right]$. Problem \eqref{im_CSI_lp} is convex with respect to $l_k$ and $p_k$, and it can be solved using standard convex optimization tools.
	
		\subsection{Optimization of $ \mathbf{W} $ while Fixing $\{\mathbf{l}, \mathbf{p}\}$ and $ \mathbf{e} $} 
		Given $ \mathbf{l} $, $ \mathbf{p} $, and $ \mathbf{\Theta} $, Problem \eqref{im_CSI} becomes a feasibility-check problem that can be reformulated as 
	\begin{equation}	\label{im_check}
		\mathrm{find} ~ \mathbf{W},\quad
			\mathrm{s.t.} ~ \eqref{secrecy rate}.
	\end{equation}

In order to achieve a better-converging solution, we transform Problem \eqref{im_check} into an optimization problem with an explicit objective function. By observing \eqref{im_CSI}, constraint \eqref{secrecy rate} should be active when the total energy consumption $ E $ is minimal. Therefore, Problem \eqref{im_check} can be expressed as 
	\begin{subequations}
		\label{maxrs}
		\begin{align}
			\underset{\mathbf{W}}{\mathrm{max}}
		~ &
	 \underset{k\in \mathcal{K}}{\sum} \left(\log_2 \left( 1 + \gamma_k \right) - \underset{\Omega_{e,k}}{\mathrm{max}}  \log_2 \left( 1 + \gamma_{e,k} \right)\right), \quad \\
			\mathrm{s.t.}~  & \eqref{secrecy rate}.
		\end{align}
	\end{subequations}
	
	Note that we only design the MUD matrix $\mathbf{W}$ in the above problem, and the term of Eve's rate $\log_2 \left( 1 + \gamma_{e,k} \right)$ is independent of $\mathbf{W}$ such that it remains constant. 
	By dropping the constant terms from the objective function, Problem \eqref{maxrs} can be simplified to the following problem,
	\begin{equation}
		\label{W}
	\underset{\mathbf{W}} {\mathrm{max}} ~ \underset{k\in \mathcal{K}}{\sum} \mathrm{log}_2\left(1+\gamma_k\right), \quad \mathrm{s.t.}~  \left\| \mathbf{w}_k\right\| =1.
	\end{equation}

	To address the non-convex objective function, we are inspired by Problem (\ref{w}) and recast the above problem as an SDP problem by introducing $ \mathbf{W}_k \triangleq \mathbf{w}_k \mathbf{w}_k^{\rm H} $ with $ \mathrm{rank}(\mathbf{W}_k) =1, \forall k \in \mathcal{K} $. We also use SDR to relax the non-convex rank-one constraint, which decouples Problem \eqref{W} into $ K $ problems,
	\begin{subequations}
		\label{opt_w}
		\begin{align}
			\underset{\mathbf{W}_k} {\mathrm{max}} ~&\phi_{a,k} (\mathbf{W}_k,\mu_{a,k}),\\
			\mathrm{s.t.}\quad \!\!\!\! & \mathrm{Tr}(\mathbf{W}_k) =1, ~\mathbf{W}_k \succeq {\mathbf 0},
		\end{align}
	\end{subequations}
	where $ \phi_{a,k} (\mathbf{W}_k,\mu_{a,k})= \mathrm{ln}\left( \sum_{j=k}^K p_j \mathrm{Tr}(\mathbf{W}_k\mathbf{H}_j)   +\sigma_a^2\right) +\mathrm{ln}(\mu_{a,k}) +1 - \mu_{a,k}\left(\sum_{j=k+1}^K p_j \mathrm{Tr}(\mathbf{W}_k \mathbf{H}_j ) + \sigma_a^2 \right) $ with $ \mathbf{H}_j = \mathbf{h}_j \mathbf{h}_j^{\rm H} $. Note that Problem \eqref{opt_w} is convex and can be solved using the interior point method.
		\subsection{Optimization of $ \mathbf{e} $ while Fixing $\{\mathbf{l}, \mathbf{p}\}$ and $ \mathbf{W} $} 
	By defining \( \mathbf{H}_k \triangleq \mathbf{H} \mathrm{diag}(\mathbf{h}_{r,k}) \in \mathbb{C}^{N_a \times M} \), and following the approach in \cite{Wu2019Intelligent}, the convergence of the solution for optimizing \( \mathbf{e} \) can be improved by introducing slack variables \( \boldsymbol{\kappa} \triangleq  \left[ \kappa_1, \cdots, \kappa_K \right] \). Here, \( \boldsymbol{\kappa} \) represents the additional security rate beyond the required offloading rate for each user \( k \). To enhance convergence, the objective is to maximize \( \boldsymbol{\kappa} \). By incorporating \( \boldsymbol{\kappa} \) into the right-hand side of constraint \eqref{im_a1}, Problem \eqref{im_check} is reformulated as	
	\begin{subequations}
		\label{opt_e}
		\begin{align}
			\underset{\boldsymbol{\kappa}, \mathbf{e}} {\mathrm{max}} \quad \!\!\!\! & \underset{k\in \mathcal{K}}{\sum} \kappa_k,  \\
			\label{rat}
			\mathrm{s.t.} \quad \!\!\!\! & \mathrm{log}_2 \left( 1 + \dfrac{p_k \left|\mathbf{w}_k^{\rm H} \left( \mathbf{h}_{a,k} + \mathbf{H}_k \mathbf{e} \right) \right|^2}{ \sum_{j=k+1}^K p_j \left|\mathbf{w}_k^{\rm H} \left( \mathbf{h}_{a,j} + \mathbf{H}_j \mathbf{e} \right) \right|^2 + \sigma_a^2} \right) \nonumber \\
			& \quad \quad- \mathrm{log}_2(1 + a_k) \ge r_k + \kappa_k, \quad\forall k \in \mathcal{K}, \\
			& \eqref{phase}, \eqref{t2}, \eqref{uncert}, \eqref{mu}.
		\end{align}
	\end{subequations}
	
	By introducing two auxiliary variables $ \rho_k $ and $ d_k $, the non-convex constraint \eqref{rat} can be reformulated as
	\begin{align}
		\label{lemma}
		\mathrm{log}_2\left(1+\dfrac{d_k}{\rho_k+\sigma_a^2}\right) + \phi_{e,k}(a_k, \mu_{e,k})& \ge r_k + \kappa_k,\\
		\label{sca1}
		p_k\left|\mathbf{w}_k^{\rm H} \left( \mathbf{h}_{a,k}+ \mathbf{H}_k\mathbf{e} \right) \right|^2 &\ge d_k,\\
		\label{sca2}
		\sum_{j=k+1}^K p_j\left|\mathbf{w}_k^{\rm H} \left( \mathbf{h}_{a,j}+ \mathbf{H}_j\mathbf{e} \right) \right|^2 &\le \rho_k.
	\end{align}
	
	Use Lemma 1 to address constraint \eqref{lemma}, and we have
	\begin{align}
		\label{1}
			\frac{1}{\mathrm{ln}2}&\left[\mathrm{ln}\left(\rho_k + d_k + \sigma_a^2\right)-\mu_{r,k} (\rho_k + \sigma_a^2)+\mathrm{ln}\mu_{r,k}+1\right]\nonumber\\
            &\quad\quad\quad \quad\quad\quad +\phi_{e,k}(a_k, \mu_{e,k}) \ge r_k + \kappa_k.
	\end{align}
	
	Note that the left-hand side of constraint \eqref{sca1} is non-concave with respect to \( \mathbf{e} \). However, it can be approximated using the SCA method. By applying the following inequality lemma for any complex \( x \),
	\begin{equation}
		\left| x \right|^2 \geq 2\mathrm{Re}\left\lbrace \tilde{x}^{\dagger}x \right\rbrace - \left| \tilde{x} \right|^2,
	\end{equation}
	where \( \tilde{x} \) is a feasible point, constraint \eqref{sca1} can be transformed into the following convex constraint,
	\begin{align}\label{2}
	 2\mathrm{Re}\left\{\left( \mathbf{h}_{a,k}^{\rm H} + \tilde{\mathbf{e}}^{\rm H} \mathbf{H}_k^{\rm H} \right) \mathbf{w}_k \mathbf{w}_k^{\rm H} \left( \mathbf{h}_{a,k} + \mathbf{H}_k \mathbf{e} \right) \right\} &-\nonumber \\
		 \left( \mathbf{h}_{a,k}^{\rm H} + \tilde{\mathbf{e}}^{\rm H} \mathbf{H}_k^{\rm H} \right) \mathbf{w}_k \mathbf{w}_k^{\rm H} \left( \mathbf{h}_{a,k} + \mathbf{H}_k \tilde{\mathbf{e}} \right)  &\geq \frac{d_k}{p_k},
	\end{align}
	where \( \tilde{\mathbf{e}} \) is the feasible point. This transformation ensures that the original non-concave constraint is approximated by a convex constraint, enabling efficient optimization.

	As a result, the problem is to optimize $ \mathbf{e} $ in the following 
	\begin{subequations}
		\label{opt_e1}
		\begin{align}
			\underset{\mathbf{e}, \mathbf{d}, \boldsymbol{\rho}, \mathbf{a}, \boldsymbol{\beta}, \boldsymbol{\kappa}}{\mathrm{max}} \quad \!\!\!\! & \underset{k\in \mathcal{K}}{\sum} \kappa_k, \\ 
			\mathrm{s.t.} ~ & \eqref{phase},  \eqref{t2}, \eqref{uncert},\eqref{mu}, \eqref{sca2}, \eqref{1}, \eqref{2},
		\end{align}
	\end{subequations}
	where $\mathbf{d} \triangleq \left[d_1,\dots,d_K\right]$ and $\boldsymbol{\rho} \triangleq \left[\rho_1,\dots,\rho_K\right]$.
	
	We note that the above problem is still non-convex due to the UMC. Here we adopt the PCC procedure to deal with the non-convex constraint. Following the PCC procedure framework, the constraint in \eqref{phase} can be first equivalently rewritten as $ 1-b_m\le \left|e_m \right|^2  \le 1+c_m$ by introducing auxiliary real vectors $ \mathbf{b} \triangleq \left[b_1,b_2,\cdots, b_M\right] $ and $\mathbf{c} \triangleq \left[c_1, c_2,\cdots, c_M\right] $ so that the feasible set of $ \left|e_m \right|  $ changes to a continuous region. The non-convex part $ 1-b_m\le \left|e_m \right|^2 $ of the transformed constraint can then be linearized by $\left|e_m \right|^2 \ge 2\mathrm{Re}\left\lbrace \tilde{e}_m^{\dagger} e_m \right\rbrace -\left|\tilde{e}_m \right|^2 \ge 1-b_m $, and then we can construct the following problem.
	\begin{subequations}
		\label{opt_e2}
		\begin{align}
			\underset{\mathbf{e}, \mathbf{d}, \boldsymbol{\rho}, \mathbf{a}, \boldsymbol{\beta}, \boldsymbol{\mu}_{\mathrm{g}}, \boldsymbol{\mu}_{\mathrm{h}}, \boldsymbol{\kappa},\mathbf{b},\mathbf{c}}{\mathrm{max}} \quad \!\!\!\! & \underset{k\in \mathcal{K}}{\sum} \kappa_k - \lambda \underset{m\in \mathcal{M}}{\sum}  \left(b_m +  c_m\right), \\ 
			\mathrm{s.t.}\quad \!\!\!\!
			& \eqref{t2}, \eqref{uncert},\eqref{mu}, \eqref{sca2}, \eqref{1}, \eqref{2},\\
			& 2\mathrm{Re}\left\lbrace \tilde{e}_m^{\dagger} e_m \right\rbrace -\left|\tilde{e}_m \right|^2 \ge 1-b_m,\\
			& \left|e_m \right|^2  \le 1+c_m,
		\end{align}
	\end{subequations}
	where $ \lambda $ is the penalty parameter to scale the impact of the penalty term. Problem \eqref{opt_e2} is an SDP problem and can be solved using the CVX tool. 
	
	The overall optimization algorithm for solving Problem (\ref{im_CSI}) is summarized as Algorithm 2. Some points are emphasized as follows: i) The maximum value $ \lambda_{max} $ is imposed to avoid numerical problems; that is, a feasible solution may not be found when the iteration converges under large values of $ \lambda^{(i)} $; ii) The stopping criterion $  \sum\nolimits_{m\in\mathcal{M}}\left( b_m +  c_m\right) \le \varepsilon_1$ guarantees that the UMC in the original Problem \eqref{opt_e1} is met for a sufficiently low { convergence tolerance} $ \varepsilon_1 $; iii) The stopping criterion $ {\left| E^{(i)} - E^{(i-1)} \right|}/{\left| E^{(i-1)} \right|} \le \varepsilon_2 $ controls the convergence of Algorithm 2.
	
	\begin{algorithm}
		\caption{Algorithm for Solving Problem (\ref{im_CSI})}\label{algorithm2}
		\hspace{0.02in}{\bf Input:} $B, C_k, T, \varsigma_k, K, \mathbf{h}_{a,k}, \overline{\mathbf{h}}_{e,k}, \mathbf{h}_{r,k}, \overline{\mathbf{G}}_k, \epsilon_{e,k}, \epsilon_{g,k}, \mathbf{H}, \varepsilon_1, \varepsilon_2$.\\
		\hspace{0.02in}{\bf Output:} $\mathbf{l}^*, \mathbf{p}^*, \mathbf{W}^*, \mathbf{e}^*$. 
		\begin{algorithmic}[1]
			\State Initialize $\mathbf{l}^{(0)}$, $\mathbf{p}^{(0)}$, $\mathbf{W}^{(0)}$, $\mathbf{e}^{(0)}$, calculate $E^{(0)}$, and set iteration number $i = 0$.
			\Repeat
			\State Solve Problem \eqref{im_CSI_lp} using LMIs with given $\mathbf{W}^{(i)}$ and $\mathbf{e}^{(i)}$, and obtain the optimal $\mathbf{l}^{(i+1)}$ and $\mathbf{p}^{(i+1)}$.
			\State Solve Problem \eqref{opt_w} with given $\mathbf{l}^{(i+1)}$, $\mathbf{p}^{(i+1)}$, and $\mathbf{e}^{(i)}$, and denote the solution after Gaussian randomization as $\mathbf{W}^{(i+1)}$.
			\State Solve Problem \eqref{opt_e2} for given $\tilde{\mathbf{e}}$ using SCA and PCC to tackle the UMC and obtain the optimal $\mathbf{e}^{(i+1)}$.
			\State Calculate $E^{(i+1)}$ according to \eqref{energy_equation} and update $i = i+1$.
			\Until{$ \sum\nolimits_{m\in\mathcal{M}}\left( b_m +  c_m\right) \le \varepsilon_1 $ or $\frac{\left| E^{(i)} - E^{(i-1)} \right|}{\left| E^{(i-1)} \right|} \le \varepsilon_2$}.
		\end{algorithmic}
	\end{algorithm}

	{ We now analyze the computational complexity of Algorithm \ref{algorithm2}, which is attributed to solving subproblems \eqref{im_CSI_lp}, \eqref{opt_w}, and \eqref{opt_e2}. Recalling to the general complexity expression \eqref{complexity}, we have \(n = 6K\), \(J = 6K+1\), \(I = 0\) for Problem \eqref{im_CSI_lp}, \(n = N_a^2\), \(J = 1\), \(I = 0\) for Problem \eqref{opt_w}, and \(n = 3M + 7K\), \(J = N_e + 9\), \(I = M\) for Problem \eqref{opt_e2}, Accordingly, we can calculate the per-iteration complexities for subproblems \eqref{im_CSI_lp}, \eqref{opt_w}, and \eqref{opt_e2} as 
		$o_{\mathbf{lp}} = \mathcal{O}\left(K^{1.5}N_e^{0.5} \left[N_e^3+K^{2}+KN_e^{2} \right] \right)$, 
		$o_{\mathbf{W}}=\mathcal{O}\left( K N_a^{6.5}\right)$, and $o_{\mathbf{e}}=\mathcal{O}\left( K\left(M^2 N_e^3 + K M N_e^3 + K^2 N_e^3 + M^3 + K M^2\right) \sqrt{M + K} \right)$, respectively.
		Therefore, the overall per-iteration complexity of solving Problem~\eqref{im_CSI} can be given by $o_{\text{tot}} = o_{\mathbf{lp}}+ o_{\mathbf{W}} + o_{\mathbf{e}}.$
	
	It is not difficult to deduce from the above analysis that the total computational complexity $o_{\text{tot}}$ of Algorithm \ref{algorithm2} grows rapidly with the RIS element number \(M\) at a dominant order of \(\mathcal{O}(M^{3.5})\), reflecting the expanding variable dimensions and constraint sizes in RIS phase shift optimization. This highlights a fundamental trade-off: while deploying more RIS elements can enhance the secrecy and energy efficiency by offering higher beamforming gain, it also incurs heavier computational burdens which may hinder real-time implementation in MEC systems. Hence, it is crucial to choose a medium-scale RIS to balance performance improvement and algorithmic tractability. For instance, for a moderate system size,  e.g., \(K=5\), \(M=64\), the proposed algorithms converge within a few seconds, which falls into a typical task offloading time-scale for IoT-MEC scenarios, thus supporting the practical operation of RIS in real-time MEC networks \cite{bai2021resource}.
}	

	\section{Numerical Results}
\begin{figure}[htbp]
		\centering	
		{\includegraphics[width=0.5\textwidth]{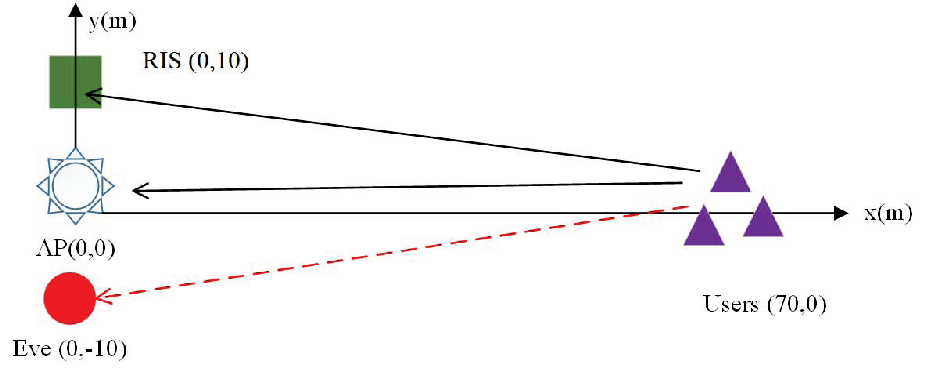}} 
		\caption{Illustration of a simulation system.} \label{simulation setup}
	\end{figure}

\begin{figure}[htbp]
		\centering	{\includegraphics[scale=0.6]{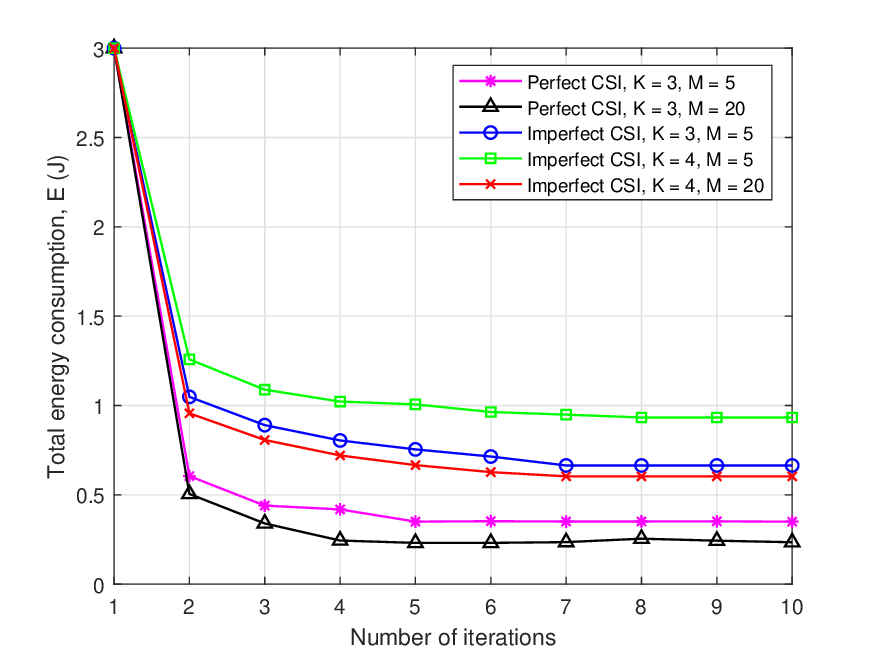}}
		\caption{Total energy consumption $E$ vs. iteration number.} \label{fig_conver}
	\end{figure}

		In this section, we provide numerical results to evaluate the performance of our proposed algorithms. The simulation system is shown in Fig. \ref{simulation setup}, in which we assume that AP, RIS, and Eve are located at (0,0), (0,10), and (0,-10), respectively, in a two-dimensional Cartesian coordinate system. $K$ users are randomly and uniformly distributed in a circle centered at (70,0) with a radius of 5 m. The channel between node $i$ and node $j$ is modeled as $h_{ij} = \sqrt{L_0 d_{ij}^{-\alpha_{ij}}} g_{ij}$, where $L_0 = -30$ dB denotes the large-scale path loss at the reference distance $d_0 = 1$ m, $d_{ij}$ is the distance in meters, $\alpha_{ij}$ is the path-loss exponent, and $g_{ij}$ is the small-scale fading component obeying Rayleigh distribution. We set the path-loss exponents of the user-AP, user-Eve, user-RIS, RIS-AP, and RIS-Eve channels as $\alpha_{u,a} = \alpha_{u,e} = 4$, $\alpha_{u,i} = 2$, $\alpha_{i,a} = 2.2$, and $\alpha_{i,e} = 2.5$, respectively. Unless otherwise stated, the other parameters are set as $N_a = 5$, $N_e = 3$, $M = 5$, $K = 3$, $\sigma_a^2 = \sigma_e^2 = -90$ dBm, $B = 1$ MHz, $T = 0.1$ s, $C_k = 1000$ cycles/bit, $\varsigma_k = 10^{-28}$, { $\lambda^{(0)} = 10$, $\lambda_{max} = 10^3$, the channel error bounds $\epsilon_{e,k} = \epsilon_{g,k} = 0.01$, and the convergence tolerance $\varepsilon=10^{-3}$ for all algorithms.} The results in Fig. \ref{fig_conver} are obtained based on a single randomly generated channel, and the results in the other figures are averaged over 50 randomly generated channels.\footnote{{Although we focus on signal design and resource allocation under given network conditions, the simulation settings can actually capture network condition variations, e.g., varying user number reflects different user densities, and choosing distinct path-loss exponents models different channel conditions.}}

  { Fig. \ref{fig_conver} presents the total energy consumption $E$ versus the iteration number of our proposed BCD algorithm for both scenarios of perfect and imperfect Eve's CSI. Just as analyzed previously, the convergence of Algorithm \ref{algorithm1} can be ensured by the convexity of each subproblem after transformation via SCA and SDR along with the monotonically deceasing objective function at each iteration, i.e., the total energy consumption. In this figure, we clearly show that $E$ monotonically decreases and converges about 5 to 10 iterations. This validates the quick convergence and effectiveness of the BCD algorithm. }

	\begin{figure}[htbp]	
		\centering
		{\includegraphics[scale=0.6]{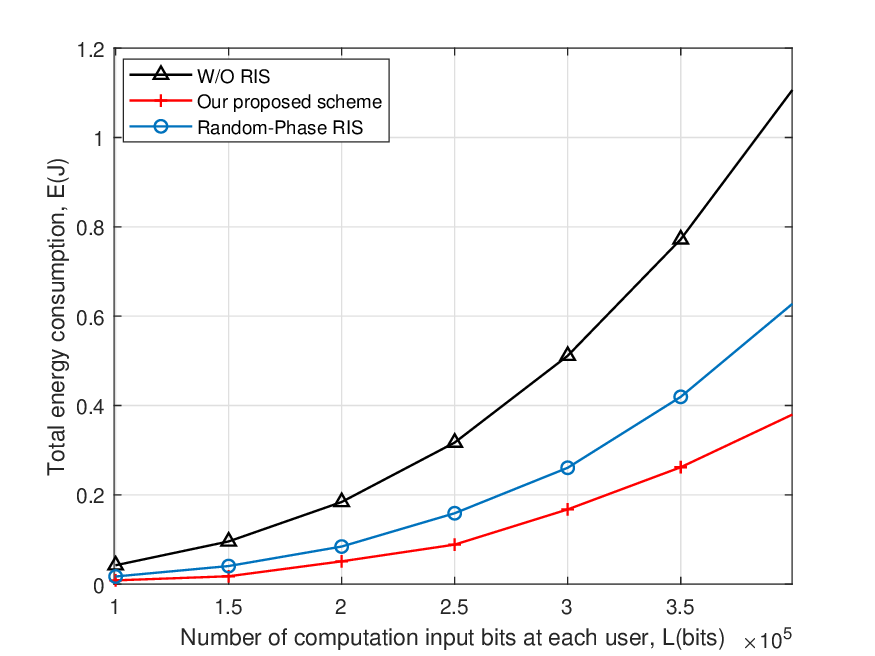}}
		\caption{Total energy consumption $E$ vs. number of computation bits at each user $L$.} \label{input bits}
	\end{figure}

	
	Fig. \ref{input bits} plots the variation of the total energy consumption $E$ with the number of computation bits at each user $L$, considering the scenario of Eve's perfect CSI. {We compare the achievable performance of our proposed scheme with two benchmark schemes, namely, (i) W/O RIS: the scheme without RIS, or equivalently $\mathbf{\Theta} = \mathrm{diag}\left(\bf{0}\right) $, and (ii) Random-Phase RIS: the scheme with random RIS phase shifts, i.e., \(\theta_m, \forall m \in \mathcal{M}\), is randomly and uniformly selected within \(\left[0,2\pi\right)\).\footnote{For the two benchmark schemes, the other parameters, i.e., $\{\mathbf{l}, \mathbf{p},\mathbf{W}$\}, still should be optimized. Note that such comparison has been commonly adopted in the context of RIS-aided PLS, e.g., in \cite{Shen2019Secrecy,Cui2019Secure,Li2022Intelligent,Lin2024Self,Dong2025Secure,Hong2020ArtificialNoiseAided}, since it directly highlights the security performance gain brought by introducing a fine-designed RIS compared to the scenarios without RIS or with an RIS of random phase shifts. On the other hand, to the best of our knowledge, there is no directly comparable algorithm tailored for our considered RIS-enabled secure and green NOMA-MEC network. Therefore, it would also be unfair if the methods which are not suitable for our system model were used as benchmark schemes, since they would unsurprisingly be inferior to ours.}
	It is as expected that $E$ monotonically increases with $L$, meaning that more energy should be consumed to deal with more computation tasks in a certain task duration.
	We observe that our proposed scheme significantly outperforms the two benchmark schemes in terms of secure energy efficiency, i.e., achieving the lowest total energy consumption for the same number of computation bits. This demonstrates the necessity and effectiveness of introducing RIS in realizing a secure and green MEC network, where the required power of secure offloading can be lower by configuring the signal propagation path intelligently with the aid of RIS. This also highlights the superiority of a joint active-passive beamforming design compared with a random-phase RIS counterpart, especially when facing a larger number of computation bits.}
	
	\begin{figure}[htbp]
		\centering	{\includegraphics[scale=0.6]{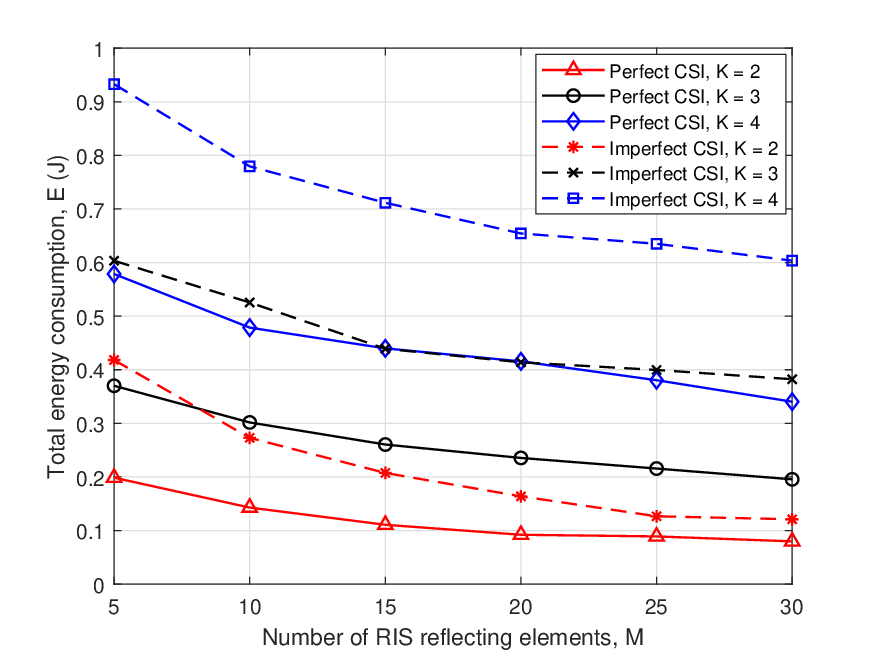}}
		\caption{Total energy consumption $E$ vs. RIS element number $M$, with $L = 3\times 10^5$ bits.} \label{fig_M}
	\end{figure} 

    Fig. \ref{fig_M} presents the total energy consumption $E$ versus the number of RIS reflecting elements $M$ for different user number $K$, considering both scenarios of Eve's perfect and imperfect CSI. {  It can be clearly seen that $E$ monotonically decreases as $M$ increases, indicating that a larger passive beamforming gain can be gained with more RIS reflecting elements, thus improving the energy efficiency of secure offloading. 
    Nevertheless, we can observe that the decrement of $E$ becomes marginal as $M$ continuously increases. 
  This phenomenon can be explained as follows: On one hand, the passive beamforming gain of RIS obeys a sublinear relationship with the number of reflecting elements due to the surface aperture constraint and the energy conservation law in passive reflection. The system will enter a channel hardening regime as more RIS element are deployed, where additional elements can only bring negligible channel enhancement. On the other hand, although a larger scale of RIS produces a larger beamforming gain, it might also increase the inter-user interference at the AP, which not only increases the decoding complexity of the NOMA offloading scheme, but also makes the SINR increase less significant gradually. Inspired by this insight, we should choose a moderate number of RIS elements (e.g., $M=20$ for $K=2$) to balance well between improving secrecy energy efficiency and lowering system overhead and computation burden during practical design. In addition, we show that due to CSI imperfection, considerable additional energy will be consumed to compensate for the secrecy rate loss, especially when more users coexist in the MEC network.}

	\begin{figure}[htbp]
		\centering	{\includegraphics[scale=0.6]{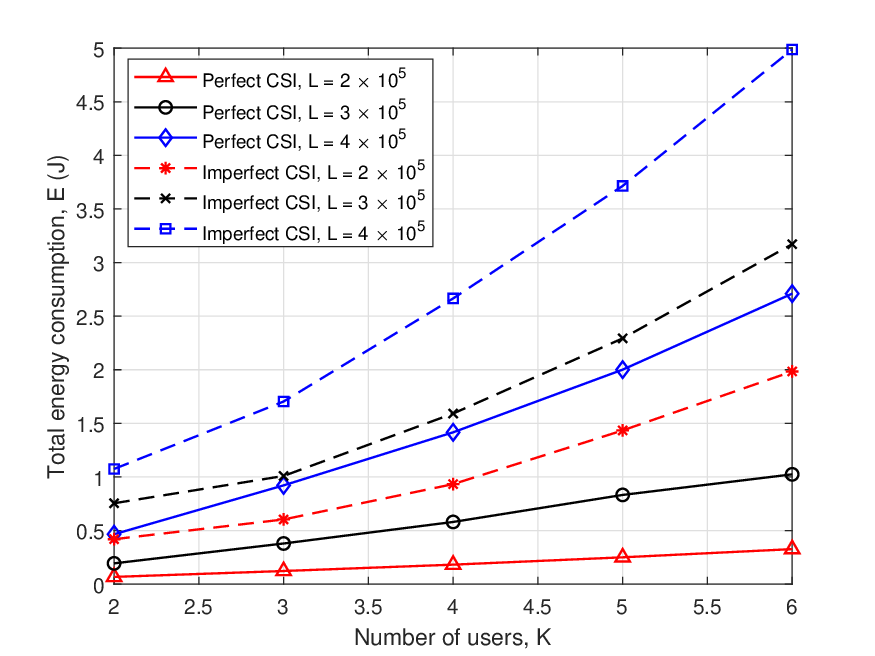}}
		\caption{Total energy consumption ($E$) vs. user number $K$.} \label{fig_K}
	\end{figure}

	 Fig. \ref{fig_K} illustrates how the total energy consumption $E$ varies with the number of users $K$ for different numbers of computation bits $L$. { As expected, $E$ monotonically increases with both $K$ and $L$ because higher transmission power is required to complete more computation tasks in a certain task duration. We also observe that the increase trend is slow particularly for a small-scale computation task, which validates the effectiveness of our proposed algorithms when applied for a large number of users. In addition, a noticeable performance gap exists between the perfect and imperfect CSI cases. The underlying reason is: RIS-assisted beamforming is highly sensitive to CSI accuracy, while CSI imperfection causes signal misalignment, then higher transmit power is required to make up for the beamforming loss. The imperfect CSI also impairs the suppression of Eve's channels, thus degrading the secrecy rate performance, which also should be compensated by consuming more energy. }

	\begin{figure}[htbp]
		\centering	{\includegraphics[scale=0.6]{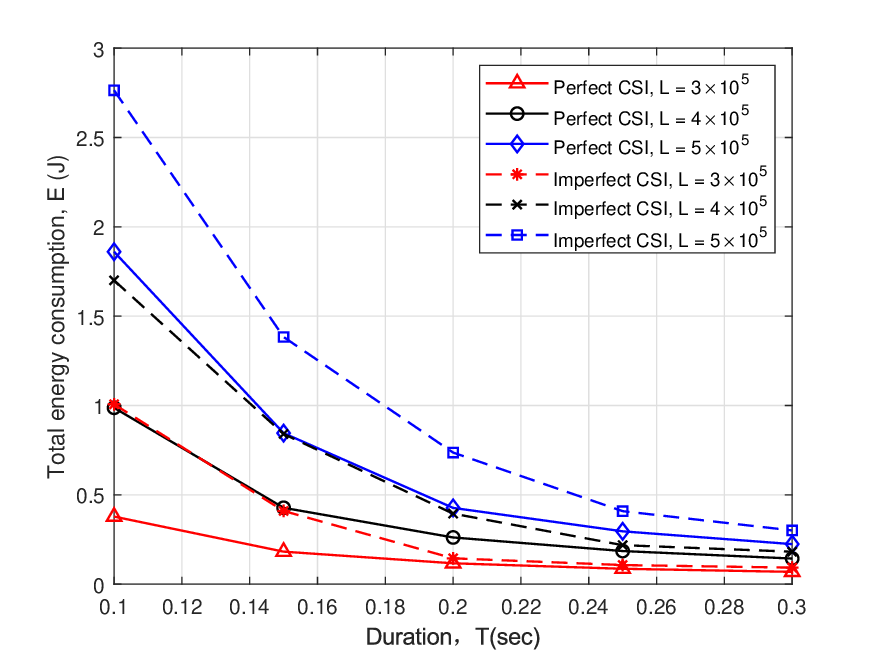}}
		\caption{Total energy consumption ($E$) vs. task duration $T$.} \label{fig_T}
	\end{figure}

    Fig. \ref{fig_T} presents the total energy consumption $E$ as a function of the task duration $T$ for different numbers of computation bits $L$. { Clearly, $E$ monotonically decreases as $T$ increases for a fixed value of $L$, indicating that less energy consumption is required if a computation task could be accomplished in a more ample duration. The reason behind is twofold: First, a lower transmit power is acceptable to support a relaxed rate requirement as the whole offloading duration increases. Meanwhile, the CPU frequency of local computation can be deceased as $f_{k,n} \propto {C_k l_k}/{T}$, leading to a square decay in local energy consumption $E_k^{\rm loc} = {\varsigma_k C_k^3 l_k^3}/{T^2}$. We also observe that the decrease trend of $E$ slows down as $T$ continuously increases. This is because, as $T$ becomes sufficiently large, all computation bits can be processed locally in time, which is more energy-efficient than offloading.}

	\begin{figure}[htbp]
		\centering	{\includegraphics[scale=0.6]{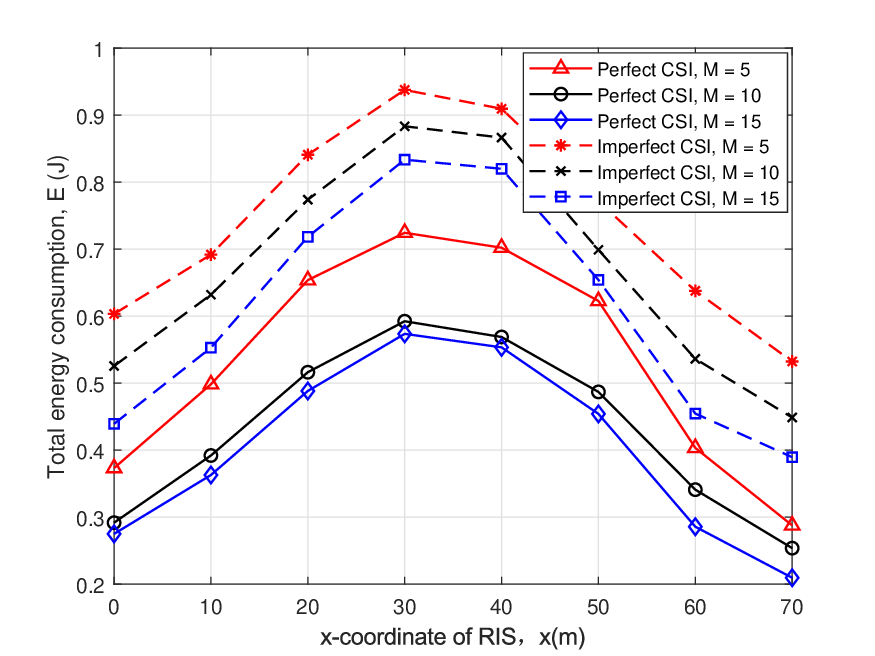}}
		\caption{Total energy consumption ($E$) vs. RIS x-coordinate $x$.} \label{fig_d}
	\end{figure}
	
    Fig. \ref{fig_d} illustrates the variation of the total energy consumption $E$ with respect to the x-coordinate of RIS $x>0$, where a smaller value of $x$ indicates that the RIS is closer to the AP as can be shown in Fig. \ref{simulation setup}. { We can see that $E$ initially increases and then decreases as $x$ increases. This is mainly because of the double path loss caused by the cascaded channel of a passive RIS, and the most severe path loss happens when the RIS is positioned approximately the midpoint between the AP and the users. Therefore, in order to minimize the total energy consumption, it is preferable to deploy the RIS near either the AP or the users. Comparing the curves of perfect CSI, we observe that the decrement of the total energy consumption is significantly large from $M=5$ to $M=10$ than that from $M=10$ to $M=15$. This implies we do not need to deploy too many reflection elements on a surface for improving the secrecy and energy efficiency, just as explained in Fig. \ref{fig_M}.}

	\begin{figure}[htbp]	
		\centering
		{\includegraphics[scale=0.6]{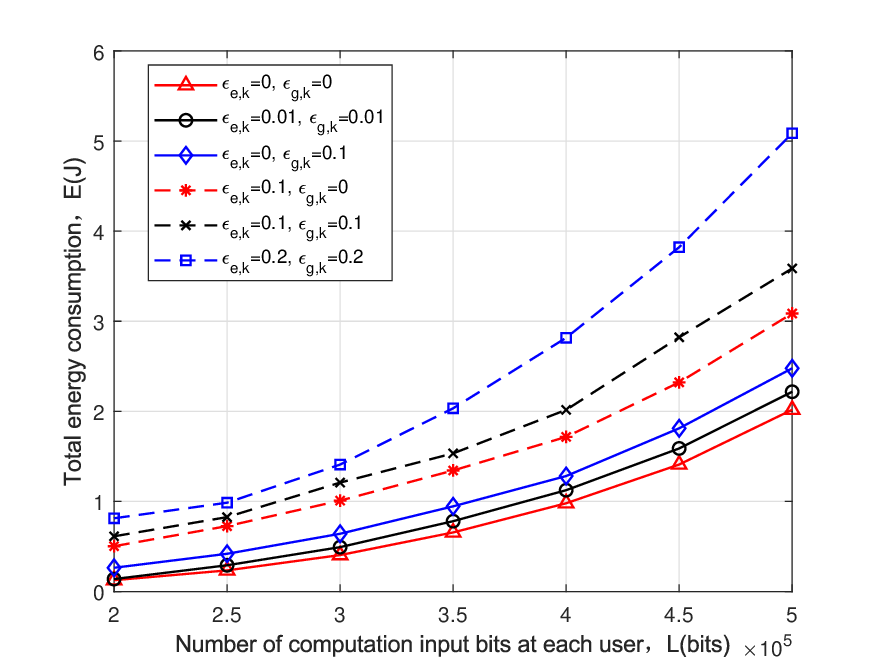}}
		\caption{Total energy consumption ($E$) vs. number of computation bits at each user $L$, with different $\epsilon_{e,k}$ and $\epsilon_{g,k}$.} \label{csi_diff}
	\end{figure}
	
	Fig. \ref{csi_diff} compares the total energy consumption $E$ as a function of the number of computation bits $L$, under different uncertainty region of channel estimation errors for the direct and reflect paths $\epsilon_{e,k}$ and $\epsilon_{g,k}$, respectively. { It is as expected that the total energy consumption is higher with channel estimation errors than that with perfect CSI (i.e., $\epsilon_{e,k}=0$ and $\epsilon_{g,k}=0$), and $E$ increases with either $\epsilon_{e,k}$ or $\epsilon_{g,k}$. This indicates that the AP should consume additional energy to compensate for the beam dispersion caused by CSI imperfection. Besides, $E$ grows faster than linearly with $L$, regardless of the level of channel estimation errors. This phenomenon can be explained as follows: on one hand, a larger $L$ not only consumes more energy for local computing but also requires higher transmission power for offloading; on the other hand, a larger $L$ intensifies the inter-user interference for the NOMA scheme, which further necessitates higher transmission power to guarantee the rate requirement. Furthermore, we can observe that the impact of channel error $\epsilon_{e,k}$ (direct path) is more significant than $\epsilon_{g,k}$ (reflect path) on the total energy consumption. The underlying reason is that, $\epsilon_{e,k}$ directly influences the estimation of leakage interference, which is critical for secure beamforming, while $\epsilon_{g,k}$ involves a cascaded channel with higher path loss and has a relatively minor effect. In other words, $\epsilon_{e,k}$ directly affects the secrecy rate optimization, further increasing energy consumption; whereas  $\epsilon_{g,k}$ primarily influences the reflected signal optimization, playing a less dominant role.}
	
	\section{Conclusion}
	{ In this paper, we investigated an RIS-enabled multi-user MEC network with hybrid local computing and edge offloading against eavesdropping attacks. We developed a joint computation resource allocation and communication parameter optimization framework and addressed the problem for minimizing the total energy consumption while guaranteeing secure offloading requirement for both perfect and imperfect CSI scenarios. We presented various numerical results to validate the proposed optimization algorithms and to demonstrate the superiority of RIS in terms of improving secrecy energy efficiency for the MEC network. We also showed that the total energy consumption monotonically increases with user number and the number of computation bits at each user, and decreases with offloading duration and RIS element number. Moreover, we revealed that it is preferable to deploy the RIS near either AP or users to reduce energy consumption, and meanwhile the RIS element number should better be moderate to strike a trade-off between system performance and overhead. }
	
	\bibliographystyle{IEEEtran}
	\bibliography{IEEEabrv,Final_version}
\end{document}